\documentclass[prd,showpacs,showkeys,preprintnumbers,nofootinbib,%
               superscriptaddress,floatfix,twocolumn,a4paper]{revtex4}

\usepackage{graphicx}
\graphicspath{{./Figures/}}
\usepackage{amssymb,amsmath,amsfonts,amsthm}
\usepackage{array,multirow}
\usepackage{epstopdf} 
\usepackage[colorlinks=true,linkcolor=black,citecolor=blue]{hyperref}
\usepackage{url}  
\newcommand{\Fig}[1]{Fig.~\ref{#1}}
\newcommand{\Tab}[1]{Table~\ref{#1}}
\newcommand{\Sec}[1]{Sect.~\ref{#1}}
\newcommand{\App}[1]{Appendix~\ref{#1}}
\newcommand{\Eq}[1]{Eq.\,\eqref{#1}}
\renewcommand{\Re}{\operatorname{\mathfrak{Re}}}
\newcommand{\Tr}{\operatorname{Tr}}

\providecommand{\abs}[1]{\lvert#1\rvert}

\begin{document}
\preprint{HU-EP-09/08, RCNP-Th09001, ADP-09-02/T680}

\title{Coulomb-gauge ghost and gluon propagators in $\boldsymbol{SU(3)}$ lattice
  Yang-Mills theory}

\author{Y.~Nakagawa} 
\affiliation{Research Center for Nuclear Physics, Osaka University,
  Ibaraki-shi, Osaka 567-0047, Japan}
\author{A.~Voigt}
\affiliation{Humboldt-Universit\"at zu Berlin, Institut f\"ur Physik,
D-12489 Berlin, Germany}
\affiliation{Max-Planck-Institut f\"ur Meteorologie, D-20146 Hamburg, Germany}
\author{E.-M.~Ilgenfritz} 
\affiliation{Humboldt-Universit\"at zu Berlin, Institut f\"ur Physik,
  D-12489 Berlin, Germany}
\affiliation{Karl-Franzens-Universit\"at Graz, Institut f\"ur Physik,
  A-8010 Graz, Austria} 
\author{M.~M\"uller-Preussker} 
\affiliation{Humboldt-Universit\"at zu Berlin, Institut f\"ur Physik,
  D-12489 Berlin, Germany}
\author{A.~Nakamura}
\affiliation{Research Institute for Information Science and Education, 
  Hiroshima University, Higashi-Hiroshima 739-8521, Japan}
\author{T.~Saito}
\affiliation{Integrated Information Center,
  Kochi University, Akenobo-cho, Kochi 780-8520, Japan}
\author{A.~Sternbeck}
\affiliation{CSSM, School of Chemistry \& Physics, The University of
  Adelaide, SA 5005, Australia} 
\author{H.~Toki}
\affiliation{Research Center for Nuclear Physics, Osaka University,
  Ibaraki-shi, Osaka 567-0047, Japan}
\date{February 25, 2009}

\begin{abstract}
We study the momentum dependence of the ghost propagator and of the
space and time components of the gluon propagator at equal time in 
pure $SU(3)$ lattice Coulomb gauge theory carrying out a 
joint analysis of data collected independently at RCNP Osaka and 
Humboldt University Berlin. We focus on the scaling behavior of these 
propagators at $\beta=5.8,\ldots,6.2$ and apply a matching technique 
to relate the data for the different lattice cutoffs. Thereby, lattice 
artifacts are found to be rather strong for both instantaneous gluon 
propagators at large momentum. As a byproduct we obtain the respective 
lattice scale dependences $a(\beta)$ for the transversal gluon and the 
ghost propagator which indeed run faster with $\beta$ than two-loop 
running, but slightly slower than what is known from the Necco-Sommer 
analysis of the heavy quark potential. The abnormal $a(\beta)$
dependence as determined from the instantaneous time-time gluon
propagator, $D_{44}$, remains a problem, though. The role of residual  
gauge-fixing influencing $D_{44}$ is discussed. 
\end{abstract}

\maketitle

\section{Introduction}
\label{sec:introduction}

Lattice investigations of the gluon and ghost propagator have become
an important topic over the last ten years after the pioneering
lattice studies in the Landau gauge appeared in the late eighties and
nineties
\cite{Mandula:1987rh,Marenzoni:1994ap,Suman:1995zg,Nakamura:1995sf,
Leinweber:1998im, Becirevic:1999uc,Mandula:1999nj} and after the
coupled solutions to the corresponding Dyson-Schwinger equations in
the deep infrared momentum region were
found~\cite{vonSmekal:1997is,vonSmekal:1997vx}. Since then, the
available amount of lattice data on these propagators has grown (see,
e.g.,
\cite{Bonnet:2000kw,Bonnet:2001uh,Bowman:2004jm,Sternbeck:2005tk,
Ilgenfritz:2006he,Sternbeck:2006cg,Sternbeck:2007ug,Bogolubsky:2007ud,
Bowman:2007du,Kamleh:2007ud,Cucchieri:2007md}) and also studies based
on functional methods have made considerable
progress~\cite{Lerche:2002ep,Zwanziger:2001kw,Pawlowski:2003hq,Fischer:2006vf}
such that in Landau gauge one is nowadays in the comfortable situation
to confront continuum results with a broad set of independent lattice
data (see, e.g., \cite{Fischer:2008uz,Sternbeck:2008mv} for recent
discussions).

In Coulomb gauge, comparably few lattice investigations of
the aforementioned propagators have been performed. For example, Langfeld
and Moyaerts~\cite{Langfeld:2004qs} as well as Cucchieri and
Zwanziger~\cite{Cucchieri:2000gu}, and very recently also Burgio, Quandt
and Reinhardt~\cite{Quandt:2007qd, Burgio:2008jr} have carried out
such computations for the gauge group $SU(2)$. In fact, the
Coulomb gauge provides an interesting alternative to the Landau gauge
since the resulting Hamiltonian approach allows to apply the
variational principle to get analytic results for the QCD vacuum wave
functional and for the 
spectrum of hadronic bound states. This approach has been mainly
pursued by the T\"ubingen group~\cite{Reinhardt:2004mm} in recent
years.  Their investigations of (truncated) systems of Dyson-Schwinger
equations in the Coulomb gauge provided various solutions in the
infrared, including also asymptotic ``conformal'' or ``scaling''
solutions that are characterized in the infrared by a singular
power-law behavior of the ghost propagator and a powerlike vanishing
transversal gluon
propagator~\cite{Schleifenbaum:2006bq,Epple:2006hv,Epple:2007ut}, very
similar to what was found in Landau gauge.

Among us, the authors from Japan have done several lattice
investigations before, concentrating on the Coulomb gauge for $SU(3)$
gauge fields. The instantaneous gluon propagators 
and the ghost propagator were computed in~\cite{Nakagawa:2007zzb}, and
correlators of incomplete Polyakov-loops were also determined. The latter 
was studied in order to interpolate between the confinement potential 
$V_{c}$ (derived from Wilson loops) and the Coulomb potential 
$V_{\mathrm{Coul}}$ (known to restrict $V_{c}$ from 
above~\cite{Zwanziger:1998ez,Zwanziger:2002sh}). Furthermore, it was 
possible in this way to extract potentials for the quark-antiquark singlet 
and octet channels, as well as for the quark-quark symmetric sextet and 
antisymmetric anti-triplet channels~\cite{Nakagawa:2006fk,Nakagawa:2008ip}. 
The eigenvalue spectrum of the Coulomb-gauge Faddeev-Popov (FP) operator 
was studied in~\cite{Greensite:2004ur,Nakagawa:2007fa}. 

Recently, some of us have also computed the gluon and ghost
propagators as well as the Coulomb potential $V_{\mathrm{Coul}}$, the
latter directly from the FP operator, however~\cite{Voigt:2007wd}. For
$V_{\mathrm{Coul}}$ very strong Gribov-copy effects were reported, and
it still remains difficult to give a final answer for the infrared
momentum limit~\cite{Voigt:2008rr}.  Independent of that, the
factorization assumption proposed in~\cite{Zwanziger:2003de} relating
$V_{\mathrm{Coul}}$ to the square of the ghost propagator was found to
be strongly violated at low momenta.

In this paper we present a joint analysis of data from Berlin and Osaka 
for the instantaneous propagators of both ghost and gluons. For the 
transverse gluon propagator as well as for the ghost propagator similar 
infrared properties as for the Landau gauge are expected also
for the Coulomb gauge. In Landau gauge, e.g., a gluon propagator vanishing
in the zero-momentum limit (or a infrared-diverging ghost dressing
function) is crucial from the point of view of the Gribov-Zwanziger 
confinement scenario~\cite{Gribov:1977wm,Zwanziger:1991gz} or the Kugo-Ojima
confinement criterion~\cite{Kugo:1979gm}. In Coulomb gauge, the 
instantaneous time-time gluon correlator should become singular and be 
related to the effective Coulomb potential.

This study provides a comprehensive set of lattice data on the
instantaneous gluon and ghost propagators in the Coulomb gauge of pure
$SU(3)$ lattice gauge theory. We discuss their momentum dependence and
analyze in detail apparent scaling violations of the space and time
components of the gluon propagator. We show that these violations can
be ameliorated if different cuts are applied on the data. In fact,
they are effectively eliminated by a matching procedure that provides
us also with the running of the lattice scale $a(\beta)$, separately
for each propagator. With the exception of the time-time 
propagator $D_{44}$, we find these runnings to be in good agreement
with other prescriptions. The behavior of the renormalization
coefficients, that are also provided by the matching procedure, is
smooth a long as $\beta \ge 6.0$.

The structure of the paper is as follows: In \Sec{sec:ensembles} we
describe the setup of our lattice simulation including details on our
gauge-fixing algorithms. \Sec{sec:propagators} introduces the relevant
lattice observables. The data for the propagators is discussed in
\Sec{sec:raw_data} where we report on obvious scaling violations for
the gluon propagators. We then use a matching procedure to relate the
propagators for different lattice cutoffs to each other and discuss
the outcome of this for the instantaneous gluon propagators and the
ghost propagator in Sects.~\ref{sec:transversal_gluon},
\ref{sec:time_time_gluon} and \ref{sec:ghost}.  We present in detail
the interplay of the matching procedure with the necessity of an
additional momentum cutoff that restricts the reliability of the data
to relatively small momenta $|pa| < \alpha$. The lattice scale
dependence $a(\beta)$, as determined thereby, is compared to what is
known from the literature in~\Sec{sec:scaling}. Finally, in
\Sec{sec:fits}, we discuss the momentum dependence of the propagators
in both the ultraviolet and infrared region. We draw our conclusions
in \Sec{sec:conclusion}.  To make the paper self-consistent, we give a
brief outline of the matching procedure for the Coulomb-gauge
propagators in \App{app:matching}. Fit and matching tables
are presented in \App{app:tables}.

\section{Lattice field ensembles and gauge fixing}
\label{sec:ensembles}

The results discussed below are based on an extensive set of quenched
gauge configurations generated in Osaka and Berlin. At both places we
employed Wilson's one-plaquette action and a standard heatbath
algorithm (including microcanonical steps) for thermalization, but
used different values of the inverse coupling $\beta$ and different 
lattice sizes $L^4$. Those, together with a couple of other useful 
parameters, are listed in~\Tab{tab:table1} that can be found in 
\App{app:tables}. 

In our analysis below we combine the data from Osaka with the data
obtained in Berlin. Both sets are nicely consistent with each other as
we checked by comparing data at $\beta=5.8$ and $6.0$.

Configurations were fixed to Coulomb gauge via maximizations of the
Coulomb gauge functional 
\begin{equation}
\label{eq:gauge_functional} 
 F_{U}[g] = \sum_{i=1}^{3} \sum_{\vec{x},t} \frac{1}{3} \Re\Tr
  U^g_i(\vec{x},t) \,,
\end{equation} 
where $U^g_i(\vec{x},t)
= g(\vec{x},t)\, U_i(\vec{x},t)\,g^{\dagger}(\vec{x}+\hat{i},t)$.  For
a fixed $U$ this was done by iteratively changing $g$ using a standard
overrelaxation (OR) algorithm in Osaka and the simulated annealing
method combined with subsequent overrelaxation (SA+OR) as in
Refs.~\cite{Voigt:2007wd,Voigt:2008rr} in Berlin. Strictly speaking,
different gauge-fixing methods may cause variations in the data of
gauge-variant observables due to the Gribov ambiguity. This is, in
particular, true for the ghost propagator at very low momentum (with
deviations up to $5\%$, see Ref.~\cite{Voigt:2008rr} for a detailed
account on that). 
 
Each maximum of $F_{U}[g]$ automatically satisfies the lattice Coulomb
gauge condition 
\begin{equation}
\label{eq:coulombgauge_cond} \nabla_i A_{i}^{c}=0 
\end{equation} 
for all color components ($c=1,\ldots,8$). Here $\nabla_i$ is the
lattice backward derivative in one of the three spatial directions
$i$, and $A_{i}^{c}$ is the lattice gluon field.  Via $A_{\mu} =
\sum_c A^c_{\mu}T^c$, with the eight generators $T^c$ of $SU(3)$ in the
fundamental representation, the gauge field components are defined in
terms of the gauge-fixed links $U_{\mu}(\vec{x},t)$ through
\begin{equation}
  \label{eq:Amu}
  A_\mu = \frac{1}{2iag_0} \Big[ U_{\mu}(\vec{x},t) -
    U^{\dagger}_{\mu}(\vec{x},t)\Big]_{\mathrm{traceless}} \, ,
\end{equation}
where $g_0$ is the bare coupling (related to $\beta=6/g^2_0$) and $a$
denotes the lattice spacing. Note that we follow the midpoint 
definition which defines $A_\mu$ at the midpoint of a link 
$U_{\mu}(\vec{x},t)$, i.e., $A_i\equiv A_i(\vec{x}+\frac{1}{2}\hat{i},t)$ 
and \mbox{$A_4\equiv A_4(\vec{x},t+\frac{1}{2})$}.

Obviously, maximization of $F_U[g]$ proceeds independently in each
time slice, as neither the Coulomb gauge functional
(\ref{eq:gauge_functional}) nor the resulting gauge condition
(\ref{eq:coulombgauge_cond}) fixes a link in temporal direction.  We
observe that the time slices of a given configuration may behave very
differently during the iterative gauge-fixing process. In fact, we
find that the number of necessary iterations may differ by a factor of
10 to 20 between the individual time slices of a given configuration. 
This obstruction reflects that a topological tunneling might happen
within one or a few subsequent time slices. In some cases there were
time slices which could not be fixed within a certain predefined
number of iterations. Then, the gauge-fixing process was repeated  
for those time slices starting from a different randomly chosen gauge  
transformation $g$ restricted to that slice, while leaving the
``well-behaved'' (already gauge-fixed) time slices untouched. In the
majority of cases, time slices did not show any recalcitrancy during
gauge fixing, though.
\begin{figure}[b]
  \includegraphics[width=0.8\linewidth]{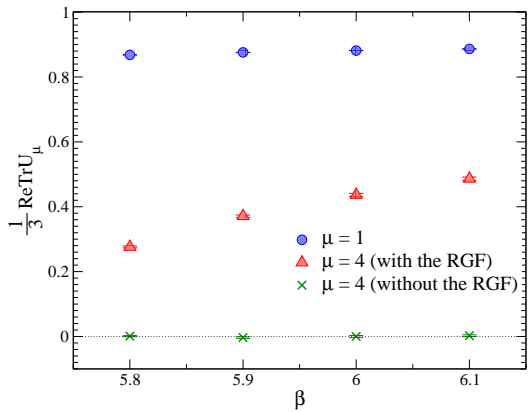}
  \caption{Mean link values $\langle\frac{1}{3}\Re\Tr U_{\mu}\rangle$
    for different directions~$\mu$ on a $18^4$ lattice. 
    Within errors the spatial links
    $\langle\Re\Tr U_i\rangle\,,~i=2,3$ (not shown) are equal to
    $\langle\Re\Tr U_1\rangle$. $\langle\frac{1}{3}\Re\Tr U_4\rangle$ 
    refers to the time-like links before and after residual gauge-fixing
    (RGF).}
  \label{fig:fig1}
\end{figure}

After all the individual time slices were maximized, the original
configuration $U$ was gauge-transformed
\begin{subequations}
  \label{eq:gaugetrafo}
\begin{align}
  \label{eq:gaugetrafo_i}
  U_{i}(\vec{x},t) &\to
  g(\vec{x},t)\,U_{i}(\vec{x},t)\, g^{\dagger}(\vec{x}+\hat{i},t)\, ,\\
  \label{eq:gaugetrafo_4}
  U_{4}(\vec{x},t) &\to
  g(\vec{x},t)\,U_{4}(\vec{x},t)\, g^{\dagger}(\vec{x},t+1)\; ,  
\end{align}
\end{subequations}
i.e., including also the time-like links. 
\begin{figure}[t]
  \includegraphics[width=0.8\linewidth]{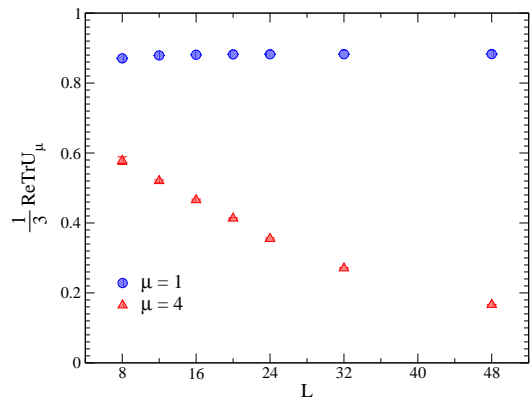}
  \caption{Mean link values $\langle\frac{1}{3}\Re\Tr U_{\mu}\rangle$
    for different directions~$\mu$ at $\beta=6.0$ as function of the
    linear lattice size $L$. Notice the strong 
    volume dependence of $\langle\frac{1}{3}\Re\Tr U_4\rangle$ after 
    residual gauge-fixing (RGF).}
  \label{fig:fig2}
\end{figure}

After having fixed the Coulomb gauge by maximizing the functional $F_U[g]$ 
there is still freedom to carry out gauge transformations $h_{t} \in SU(3)$ 
which only depend on time. One way to fix this residual gauge freedom is to 
maximize the functional
\begin{equation}
  \label{eq:residual} 
  \widetilde{F}_{U}[h] = \sum_{t} \frac{1}{3} \Re\Tr
  \left\{h(t)\left[\sum_{\vec{x}} U_{4}(\vec{x},t)\right]
    h^{\dagger}(t+1)\right\}
\end{equation}
where $U$ is the Coulomb-gauge transformed
configuration~(Eqs.~\ref{eq:gaugetrafo}). Links in time direction are
finally gauge-transformed under $h(t)$ as
\begin{subequations}
  \label{eq:residul_GT}
\begin{align}
  \label{eq:residul_GT_4}
  U_{4}(\vec{x},t) \to h(t)U_{4}(\vec{x},t) h^{\dagger}(t+1)
\intertext{whereas spatial links are transformed as}
  \label{eq:residul_GT_i}
  U_{i}(\vec{x},t) \to h(t)U_{i}(\vec{x},t)h^{\dagger}(t)\, , 
\end{align}
\end{subequations}
which preserves the Coulomb gauge. Equal-time observables involving
only spatial links, i.e., the transversal gluon or the ghost
propagator, are not affected by this residual gauge freedom.

If the transversal or the time-time gluon propagator was to be defined
for non-equal time, the residual gauge would have to be fixed as well.
For equal times, however, it is not clear to what extent, if at all,
the results for the instantaneous time-time gluon propagator
$D_{44}^{ab}(\vec{x})$ would change if the remaining gauge freedom was
fixed. We will check this in \Sec{sec:time_time_gluon} by comparing
data for $D_{44}$ for fixed residual gauge freedom (Berlin data) to
that where this freedom was left unfixed (Osaka data).

In \Fig{fig:fig1} we show the average trace $\langle\frac{1}{3}\Re\Tr
U_{j}\rangle$ for spatial links after Coulomb gauge fixing (invariant
under residual gauge fixing) and also the average
$\langle\frac{1}{3}\Re\Tr U_{4}\rangle$ before and after residual
gauge fixing as function of $\beta$. The data is taken for a $18^4$ 
lattice. Without residual gauge-fixing the average trace of
time-like links vanishes, whereas after residual gauge-fixing the
expectation value is finite and increases with increasing $\beta$. 
\Fig{fig:fig2} demonstrates that the average 
$\langle\frac{1}{3}\Re\Tr U_{4}\rangle$ after residual gauge-fixing
(shown for $\beta=6.0$) steeply decreases with increasing lattice volume, 
in contrast to $\langle\frac{1}{3}\Re\Tr U_{j}\rangle$ for spatial links.

In~\Sec{sec:time_time_gluon} we will demonstrate that the difference of
$D_{44}$ measured with and without residual gauge-fixing can be
completely accounted for by a multiplicative, momentum-independent
rescaling, {\it e.g.}, by normalizing the matched $D_{44}$ propagator
from both versions at some reference scale $p = \mu$.  
Apart from this, the residual gauge fixing has an impact only on the
value of the propagator at zero momentum but this is not of importance
for our present study.

\section{Coulomb-gauge propagators on the lattice}   
\label{sec:propagators}

The space and time components of the gluon field evaluated in momentum
space enter the bare instantaneous transversal and time-time gluon
propagator as the Monte Carlo correlators
\begin{subequations}
\label{eq:equaltimegluon}
\begin{align}
  \label{eq:equaltimegluon_ij}
  D^{ab}_{ij}(\vec{k}) &= \left\langle \tilde{A}^a_i({\vec k})
    \tilde{A}^b_j({-\vec k}) \right\rangle\, ,\\
  \label{eq:equaltimegluon_44}
  D^{ab}_{44}(\vec{k}) &= \left\langle \tilde{A}^a_4({\vec k})
    \tilde{A}^b_4({-\vec k}) \right\rangle\, .
\end{align}
\end{subequations}
Here $\tilde{A}_i$ and $\tilde{A}_4$ denote the spatial Fourier
transforms of the lattice gluon fields at a fixed time $t$ with
integer momenta $k_i\in \left(-L/2,L/2\right]$. An average over all
time slices is understood. $D^{ab}_{ij}$ is diagonal in color 
space and transverse in momentum space. On the lattice it takes  
the form
\begin{equation}
  \label{eq:gluontensorstructure}
  D^{ab}_{ij}(\vec{k}) = \delta^{ab} \left( \delta_{ij} -
    \frac{p_i(\vec{k}) p_j(\vec{k})}{p^2} \right) D_{\rm tr}(p)
\end{equation}
with
\begin{equation}
  \label{eq:momenta_i}
  p_i(\vec{k}) \equiv \frac{2}{a}
  \sin\left(\frac{\pi k_i}{L}\right)\,.
\end{equation}
This is simply due to the lattice Coulomb gauge condition which
in momentum space translates into 
\begin{equation}
  \sum_{i=1}^3 p_i(\vec{k})\tilde{A}_i(\vec{k}) = 0\, 
\end{equation}
for all ${\vec k}$. 
In the following we use $p\equiv\big\lvert\,\vec{p}(\vec{k})\,\big\rvert$ 
to simplify the notation wherever applicable. 

When analyzing data on $D^{ab}_{ij}$ it is natural to associate the
physical momentum with $p$. Lattice results then reproduce the
continuum tensor structure of $D^{ab}_{ij}$.  Deviations from its
tree-level form are described by the dimensionless dressing function
$Z_{\mathrm{tr}}(p)$, defined by
\begin{equation}
  Z_{\mathrm{tr}} = p\,D_{\mathrm{tr}}(p)\, .
\end{equation}

Analogously, the time-time gluon propagator $D^{ab}_{44}$ may be 
presented in the form of either $D_{44}(p)$ or~$Z_{44}(p)$. Both 
are related to the full propagator through 
\begin{equation}
  \label{eq:instantenousgluon}
  D^{ab}_{44}(\vec{k}) = \delta^{ab}\,D_{44}(p) =
  \delta^{ab}\frac{Z_{44}(p)}{p} \, . 
\end{equation}
Depending on the particular focus, data below is presented in either
one or the other form.

The ghost propagator is defined as the expectation value of the
inverse Faddeev-Popov (FP) operator $M$ 
\begin{equation}
   \label{eq:ghostpropagator1} 
   \left\langle\left(M^{-1}\right)^{ab}_{xx'}\right\rangle =
   \delta_{tt'} \delta^{ab} G(\vec{x}-\vec{x}')
\end{equation}
at a fixed time $t=t'$ (subsequently averaged over all time-slices).
The FP operator is local in time and, by virtue of the chosen Coulomb
gauge functional (\ref{eq:gauge_functional}), has on the lattice the
dimensionless form
\begin{widetext}
\begin{align}\nonumber
  M^{ab}_{xx'} = \delta_{tt'}\sum_{i=1}^{3}
  \Re\Tr\Big[\left\{T^a\!,T^b\right\}&\left(U_{i}(\vec{x},t) +
  U_{i}(\vec{x}-\hat{i},t)\right) \delta_{\vec{x}\vec{x}'}\\
  &-2\,T^bT^a\,
  U_{i}(\vec{x},t)\,\delta_{\vec{x}+\hat{i},\vec{x}'}
  - 2\,T^aT^b\,
  U_{i}(\vec{x}-\hat{i},t)\,\delta_{\vec{x}-\hat{i},\vec{x}'}\Big]
  \label{eq:FP_operator}
\end{align}
\end{widetext} 
where $\hat{i}$ is a unit vector in spatial direction, $x\equiv(\vec{x},t)$
and $T^a$ is a generator of $SU(3)$ in the fundamental representation. 

We are particularly interested in the momentum dependence of the ghost
dressing function 
\begin{equation}
  J(p^2)= \left({\vec p}({\vec k})\right)^2~G({\vec p}({\vec k})) \, ,
\end{equation}
where
\begin{equation}
   \label{eq:ghostpropagator2} 
   G(\vec{p}({\vec k})) = 
   \frac{a^2}{8 L^3}
   \sum_{c,\vec{x},\vec{y}} e^{2\pi i\vec{k}\cdot(\vec{x}-\vec{y})/L}
   \left\langle\left(M^{-1}\right)^{cc}_{\vec{x},t;\vec{y},t}\right\rangle\, .
\end{equation}
Working in momentum space, it is convenient to invert $M$
for a selection of momenta and colors $c$ forming right-hand side
plane-wave sources $\xi^c(k)=\delta^{ac}e^{2\pi ik\cdot x/L}$ with
$k=(\vec{k},0)\neq 0$.  We use a preconditioned conjugate-gradient
algorithm described in Ref.~\cite{Sternbeck:2005tk}, adapted to
Coulomb gauge, to accelerate the inversion of $M$. Alternatively, we
could have used a selection of point sources $\xi$ and
Fourier-transformed the vectors $[M^{-1}\xi](x)$ providing an
estimator for the ghost propagator at once for all momenta, however
with less statistical accuracy. The plane-wave method automatically
ensures that $J(p^2)$ is averaged over all time slices. Moreover,
translational invariance is exploited to improve the estimator. Note
that $M$ cannot be inverted for $\vec{k}=0$ due to its eight trivial
(constant) zero eigenmodes.

Multiplicative renormalizability is a well-established property of the
gluon and ghost propagators in Landau gauge. In Coulomb gauge, to our
knowledge, this has been proven yet only up to one-loop by Watson and
Reinhardt quite recently~\cite{Watson:2007vc}. Their result for the
bare dressing functions obtained in the 4D momentum space, formally
translated from dimensional to lattice regularization 
looks as follows (omitting possible lattice corrections) 
\begin{align}
 \nonumber 
  Z^L_{\mathrm{tr}} &= 
    1 + g_0^2\,C_{\mathrm{tr}} \left[\log\left(a^2(x+y)\right) +
      H_{\mathrm{tr}}(\zeta)\right] + O(g_0^4), \\
 \nonumber
  Z^L_{44} &= 
    1 + g_0^2\,C_{44} \left[\log\left(a^2(x+y)\right) +
      H_{44}(\zeta)\right] + O(g_0^4), \\ 
  J^L &= 1 + g_0^2\,C_{J} \left[\log(a^2 y) + H_{J}\right] +
    O(g_0^4),
 \label{perturbative_Zs}
\end{align}
where the momentum variables are $x=p_4^{\,2}, y=\vec{p}^{\,2}, \zeta=x/y$, 
and the lattice cutoff is $a^{-1}$. 
$C_{\mathrm{tr}}, C_{44}, C_{J}$ and $H_{J}$ denote constants. 
Note the non-trivial dependence on $\zeta$ for both the Coulomb-gauge 
gluon dressing functions. When multiplicatively renormalizing the
dressing functions, e.g., in a momentum subtraction (MOM) scheme at some 
scale $\mu$, this dependence has to be carefully taken 
into account (see also \cite{Burgio:2008jr}), in particular for
equal-time correlators which according to \Eq{perturbative_Zs} require
an integration over $p_4$ or $\zeta$. Of course, $a^{-1}\gg \mu$ has
to be ensured for that which, admittedly, is very difficult to achieve
in nowadays lattice computations. Even more, lattice computations are
typically carried out at several values of $\beta$, i.e., at different
cutoff-values. 
In general one should expect, that the corresponding dressing functions
at different lattice spacings, say $a$ and $\overline{a}$, are related
to each other by a finite renormalization of the $Z$-factors which
will obviously depend only on the ratio $\overline{a}/a$ and not on
the momenta. This will then hold also for the dressing functions
\Eq{perturbative_Zs} and correspondingly also for the equal-time
correlators.

In the case of Landau gauge those $Z$-factors turned out to be close
to unity for the gluon and ghost dressing functions at similar values
of $\beta$. Therefore, it is more or less sufficient for them to
express the various lattice spacings by a unique physical scale, e.g.,
via the Sommer-scale parameter $r_0=0.5~\text{fm}$ and the
interpolation formula of Necco and Sommer~\cite{Necco:2001xg}
\begin{align}
  \ln(a/r_0) &= -1.6804 - 1.7331(\beta-6.0) \nonumber \\ 
        \hspace{-2em}     &+ 0.7849(\beta-6.0)^2 - 0.4428(\beta-6.0)^3
\label{eq:necco-sommer-interpolation}
\end{align}
obtained from the lattice analysis of the static quark-antiquark potential
and applicable in the range $5.7 \le \beta \le 6.92$. 
The remaining lattice artifacts were sufficiently dealt with by applying
cone and cylinder cuts to the momenta~\cite{Leinweber:1998uu}. While a cone 
cut addresses finite-volume effects, the cylinder cut is an easy and
effective method to reduce artifacts due to the broken rotational
symmetry. We shall apply both these cuts also to our data shown below.
However, in what follows we will demonstrate that the approach, even if
sufficient for Landau gauge, is not quite enough for the case of Coulomb
gauge. In fact, beside applying the usual cone and cylinder cuts one
has to restrict momentum components to also satisfy $a p_i \le \alpha
< 2$ and to apply non-trivial finite renormalizations between
the different cutoff values. 

Also, we shall not use \emph{a priori}  
the Necco-Sommer scaling relation [\Eq{eq:necco-sommer-interpolation}]
but instead find the specific scaling behavior for each of the
propagators defined above and present their data in terms of the
finest available lattice scale at either $\beta=6.20$ or $6.10$. For
this we employ the matching procedure of Ref.~\cite{Leinweber:1998uu}
adapted here to Coulomb gauge. A detailed outline of
this method applied to our propagators is given in \App{app:matching}.

\section{Discretization errors of the bare lattice data}
\label{sec:raw_data}

\begin{figure*}
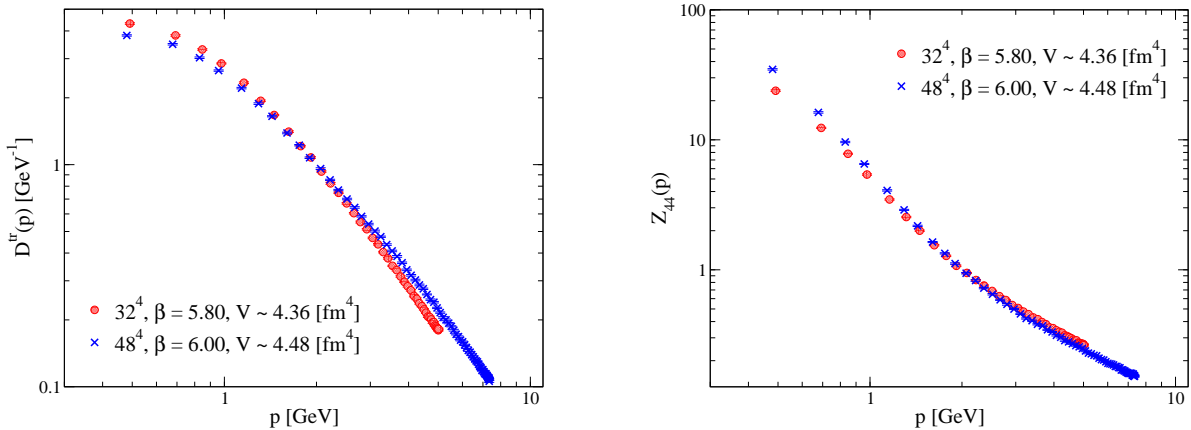

\vspace{0.2cm}
\centering
\mbox{
   \includegraphics[width=0.4\linewidth]{Dij_PhysV_unmatch}
   \qquad\qquad
   \includegraphics[width=0.4\linewidth]{Z44_PhysV_unmatch}}
\caption{The instantaneous transverse gluon propagator (left) and the
  dressing function of the instantaneous time-time gluon propagator
  (right) using the Necco-Sommer scaling relation and normalized at
  $\mu= 2~\textrm{GeV}$. The data refers to approximately equal
  physical volumes and has been cylinder and cone cut. Data was
  produced at HU Berlin.} 
\label{fig:fig3}
\end{figure*}

\begin{figure}[h]
\centering
\includegraphics[width=0.8\linewidth]{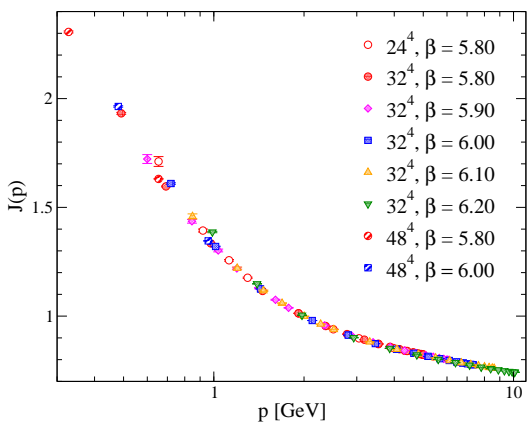}
\caption{The ghost dressing function using the Necco-Sommer 
  scaling relation and normalized at $\mu=2 \mathrm{~GeV}$.
  The data has been cylinder and cone cut. Data at $\beta=5.8,6.0,6.2$
  was collected at HU Berlin, data at $\beta=5.9, 6.1$ at RCNP Osaka.}
\label{fig:fig4}
\end{figure}

We start our discussion by revisiting the strong scaling violations we 
reported for the transversal and the time-time gluon propagator
in~\cite{Nakagawa:2007zzb,Voigt:2007wd}. There, we used the 
interpolation formula \Eq{eq:necco-sommer-interpolation}  
to assign physical units to the lattice momenta and applied 
a multiplicative normalization at $\mu=2 \mathrm{~GeV}$ for all 
values of $\beta$. This procedure, however, leads to serious
disretization errors for both the instantaneous transversal and the
time-time gluon propagator (see \Fig{fig:fig3}), whereas the ghost
propagator looks much more satisfactory in this respect  (see
\Fig{fig:fig4}). 


Challenged by these scaling violations, in the next sections we shall
perform a matching procedure that merges the data for different
$\beta$ into one bare lattice propagator associated with the highest
available lattice cutoff. 
\begin{figure*}
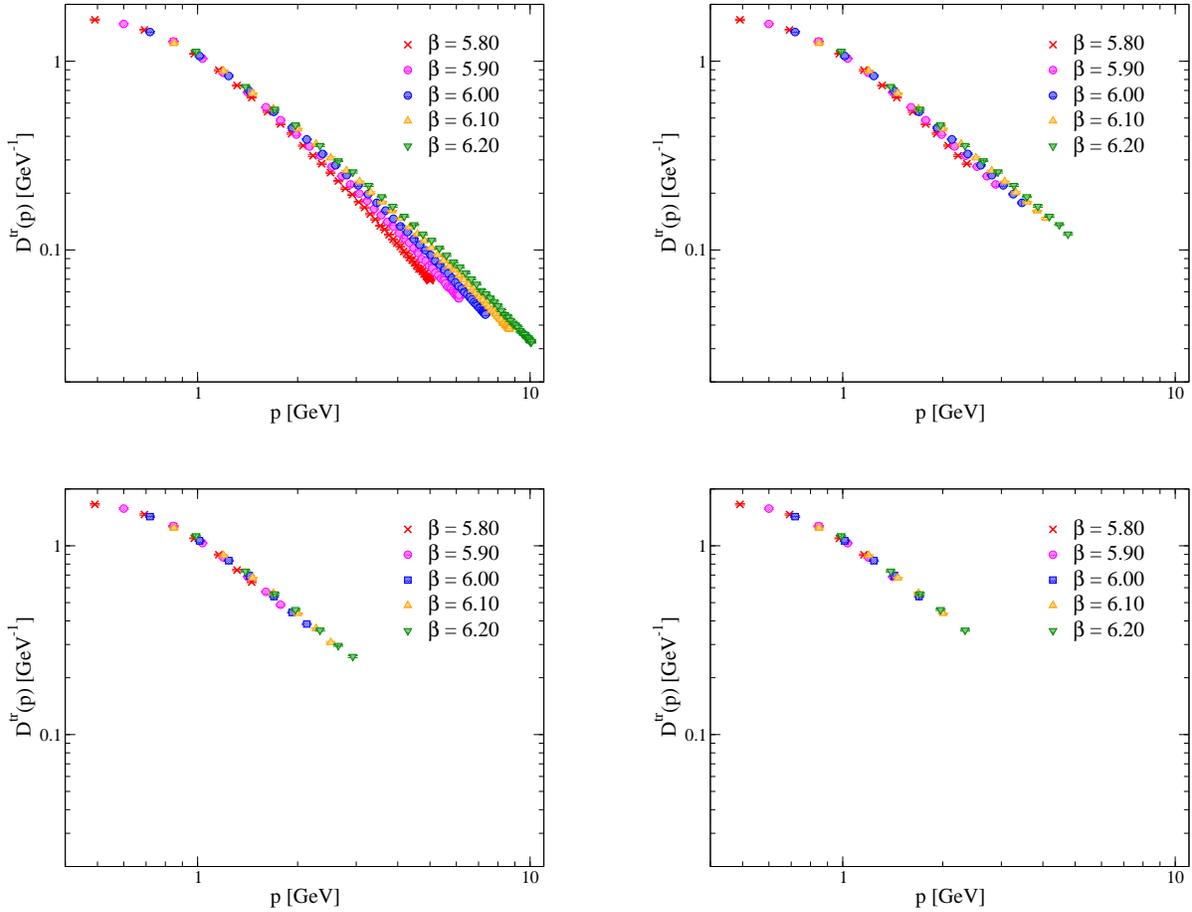

\vspace{0.2cm}
\centering
  \mbox{\includegraphics[width=0.4\linewidth]{Dij_32_Phys_Unren_Nofurthercut}
 \qquad\qquad
  \includegraphics[width=0.4\linewidth]{Dij_32_Phys_Unren_100}}\\*[0.8cm]  
  \mbox{\includegraphics[width=0.4\linewidth]{Dij_32_Phys_Unren_060}
    \qquad\qquad
    \includegraphics[width=0.4\linewidth]{Dij_32_Phys_Unren_050}}
  \caption{Effect of the $\alpha$-cut on the bare lattice data of the
    instantaneous transverse gluon propagator. The data (all for
    a $32^4$ lattice) is presented imposing different $\alpha$-cuts:
    no $\alpha$-cut (top left), $\abs{p_i a} \le 1$ (top right),
    $\abs{p_i a} \le 0.6$ (bottom left), $\abs{p_i a} \le 0.5$ (bottom
    right). Cylinder and cone cuts have been applied as
    before. Necco-Sommer scaling (with $r_0=0.5~\textrm{fm}$) has been
    used to get $a(\beta)$ and physical momenta $p$. The data for
    $\beta=5.8$, 6.0, 6.2 (5.9, 6.1) was obtained at HU Berlin
    (RCNP Osaka).}
\label{fig:fig5}
\end{figure*}

In a first step, however, we consider here the scaling violations and
argue them to indicate that the admissible range of lattice momenta
needs to be restricted even further than what the cylinder and cone
cuts would do. For this, we introduce a new momentum cut that will be
applied in addition to those two cuts. Basically, not the full
Brillouin zone should be eligible when analyzing the corresponding
propagator data, but only that at momenta (\Eq{eq:momenta_i}) whose
components are restricted to $\abs{p_ia}\leq\alpha < 2$.  For the sake
of brevity we will refer to this cut as the ``$\alpha$-cut'' in what 
follows. 

\Fig{fig:fig5} illustrates the effect of the $\alpha$-cut on the
instantaneous transversal gluon propagator. Note that in this figure (as  
in~\cite{Nakagawa:2007zzb,Voigt:2007wd}) we have used the Necco-Sommer
formula \eqref{eq:necco-sommer-interpolation} (and
$r_0=0.5~\textrm{fm}$) to assign physical units to momenta and
propagator. Obviously, when decreasing $\alpha$ less and less data
points survive this cut but those that do show a much better overlap 
than before (see, in particular, the lower panels of \Fig{fig:fig5}). 

\section{Matching the transversal gluon propagator}
\label{sec:transversal_gluon}

Still, the disagreement between data from different $\beta$ does not
 completely disappear. Therefore, in a next step, we relax the
\emph{a priori} universal $a(\beta)$ dependence (e.g., that according
to \Eq{eq:necco-sommer-interpolation}) and apply the matching
procedure of Ref.~\cite{Leinweber:1998uu} 
as explained in \App{app:matching}. It provides us with multiplicative
renormalization factors depending on the ratios of the lattice spacings
and with the specific dependence of the lattice spacing
$a=a(\beta)$ separately for each propagator. 

\begin{figure*}
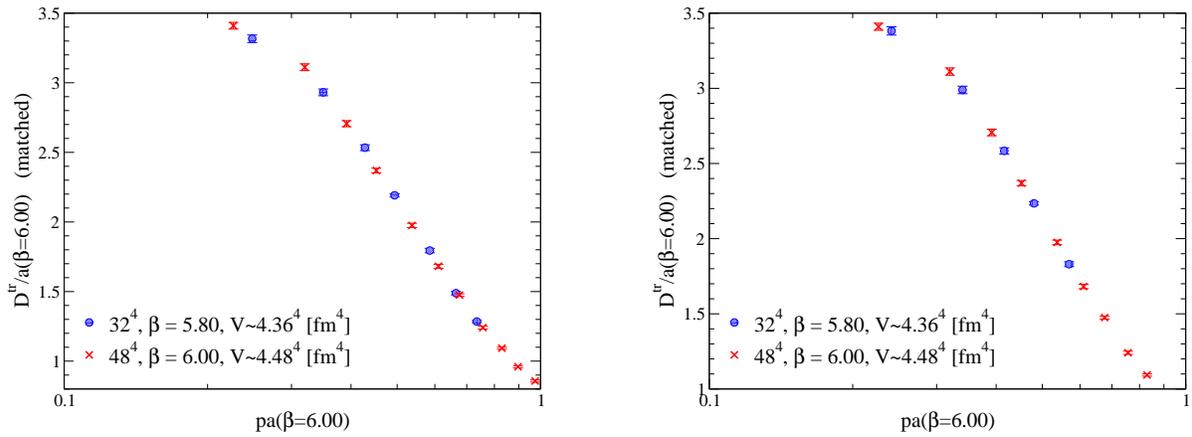

\vspace{0.2cm}
  \centering
  \mbox{\includegraphics[width=0.4\linewidth]{Dij_PhysV_matched_060}
    \qquad\qquad
    \includegraphics[width=0.4\linewidth]{Dij_PhysV_matched_050}}
  \caption{The instantaneous transverse gluon propagator obtained
   by matching data from two lattices with approximately equal
   physical volume. Besides cylinder and cone cuts, an
   $\alpha$-cut has been imposed with $\alpha = \abs{p_ia} \le 0.6$ (left)
   and $\alpha=\abs{p_i a} \le 0.5$ (right). Data collected at HU
   Berlin.} 
\label{fig:fig6} 
\end{figure*}

We start with the instantaneous transversal gluon propagator and first
match data obtained on two lattices with approximately the same
physical volume, i.e., data on a $L^4=32^4$, $\beta=5.8$ lattice with
data on a $L^4=48^4$, $\beta=6.0$ lattice. Besides the cone and
cylinder cuts we apply two different $\alpha$-cuts (with $\alpha=0.6$
and $\alpha=0.5$) before performing the matching procedure. Our aim is
to compare the influence of the $\alpha$-cut on the quality of
matching. The result, with $L^4=32^4$, $\beta=5.8$ being the coarse
and $L^4=48^4$, $\beta=6.0$ being the fine lattice, is shown
in~\Fig{fig:fig6} for both $\alpha$-cuts.

We obtain good matching of both data sets with a better result for
$\alpha=0.5$ (see the $\chi^2/dof$ listed in~\Tab{tab:table2}). There
is hardly any difference between the best result of the matching
procedure on one hand and directly imposing the Necco-Sommer scaling
relation on the other (\Fig{fig:fig5}). Indeed, our matching procedure 
nearly reproduces the lattice-spacing ratios as given through
\Eq{eq:necco-sommer-interpolation} (see~\Tab{tab:table2}) in
\App{app:tables}. 

Next we extend the matching to {\it all} values of $\beta=5.8 \ldots 6.2$
using data obtained on $32^4$ and $48^4$ lattices. Since $\beta=6.2$
has the finest lattice spacing, the matching is performed
between data at $\beta=6.2$ (setting the reference scale)
and data at all other $\beta=5.8 \ldots 6.1$. We also compare the
result for four different $\alpha$-cuts. The results are summarized
in~\Tab{tab:table3}. 

We not only find the ratios of lattice spacings to rise monotonously
upon decreasing $\beta$, but also the ratios of the renormalization
constants to be about the same (somewhat below $1.0$), i.e.,
almost independent on $\beta$.

\begin{figure}[t]
  \includegraphics[width=0.8\linewidth]{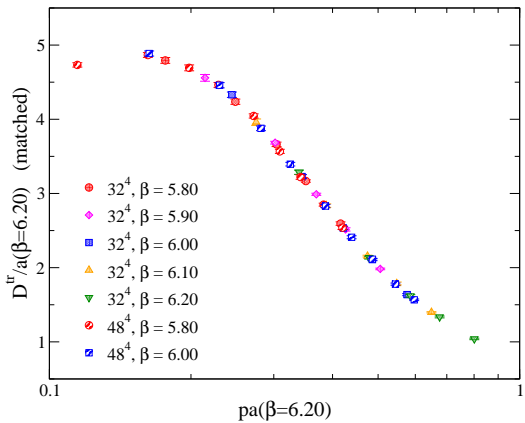}
  \caption{The instantaneous transverse gluon propagator obtained
   by matching all data from $32^4$ and
   $48^4$ lattices, including data from both Osaka and Berlin
   collected at five different $\beta$ values. The result is 
   shown for a fixed $\alpha$-cut with $\abs{p_i a} \le 0.5$. 
   Data are cylinder and cone cut.} 
\label{fig:fig7}
\end{figure}

As a general rule, smaller values for $\alpha$ result in lower
$\chi^2/dof$ values and hence better matching. We also see that the
matching procedure nearly always results in a lattice-spacing ratio
smaller than that given through \Eq{eq:necco-sommer-interpolation},
although the discrepance decreases with $\alpha$ taken smaller. As
shown in \Fig{fig:fig7} for the best $\alpha$-cut ($\alpha=0.5$),
we achieve a virtually perfect matching of the instantaneous
transversal gluon propagator over all data obtained at $\beta=5.8
\ldots 6.2$. Comparing with the $\chi^2/dof$ for $\alpha=1.0$, we
conclude that applying the $\alpha$-cut is essential to achieve a good
overlap of the data. 
Our combined final result indicates a flattening of the propagator in 
the infrared region which is worth to be explored further. We expect a
tendency to show a plateau as recently seen in the Landau 
gauge case \cite{Bogolubsky:2009dc}, which excludes a vanishing 
gluon propagator in the infrared limit.

\section{Matching the time-time gluon propagator}
\label{sec:time_time_gluon}

\begin{figure*}
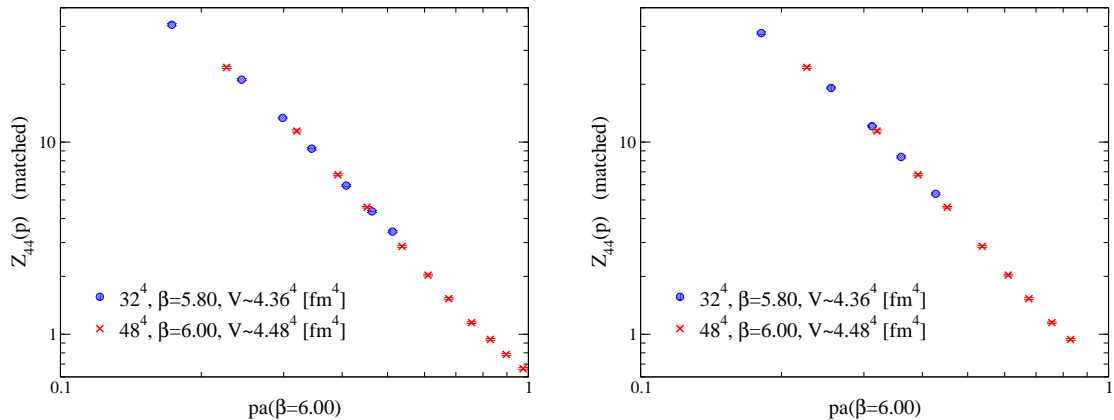

\vspace{0.2cm}
  \centering
  \mbox{\includegraphics[height=5.5cm]{Z44_PhysV_matched_060}
    \qquad
    \includegraphics[height=5.5cm]{Z44_PhysV_matched_050}}
  \caption{The instantaneous time-time gluon dressing function
   after matching Berlin data from two lattices of 
   approximately equal physical volume and applying cylinder 
   and cone cuts. Two $\alpha$-cuts are compared: 
   $\abs{p_i a} \le 0.6$ (left) and $\abs{p_i a} \le 0.5$ (right).}
\label{fig:fig8}
\end{figure*}

We now apply the matching procedure to the instantaneous time-time
gluon propagator $D_{44}$.  As for the transversal propagator
$D_{tr}$, we first match data obtained on two lattices with
approximately the same physical volume, i.e.,  data for
$L^4=32^4$, $\beta=5.8$ with data for $L^4=48^4$ and $\beta=6.0$,
respectively. We also compare the effect of two different $\alpha$-cuts
($\alpha=0.6$ and $\alpha=0.5$) in addition to the usual cylinder
and cone cuts.

Whereas the matching seems to work reliably as shown in \Fig{fig:fig8}
and as demonstrated in \Tab{tab:table4} by the low $\chi^2/dof$ value
for $\alpha=0.5$, the obtained lattice spacing ratio is now
significantly larger than predicted by the Necco-Sommer scaling
relation (see~\Tab{tab:table4}). This is in striking contrast to what
we have observed in the case of the transversal gluon propagator.

We now merge \emph{all} data for $D_{44}$ in the interval
$\beta=5.8 \ldots 6.2$ obtained on $32^4$ and $48^4$
lattices. However, since the instantaneous time-time gluon propagator 
is more sensitive to Gribov copies \cite{Voigt:2007wd} and since we
have employed different gauge fixing procedures at HU Berlin and at
RNCP Osaka, we first match the corresponding data sets separately. The
resulting fit parameters are summarized in~\Tab{tab:table5}
(Osaka) and~\Tab{tab:table6} (Berlin).

Matching the Osaka data one finds that the ratios $R_a$ of lattice
spacings rise monotonously upon decreasing $\beta$ but much stronger
than in \Eq{eq:necco-sommer-interpolation}. The ratio of
renormalization constants is still compatible with unity for
$\beta=6.0$ if compared to $\beta=6.1$ (providing the reference
scale), but it decreases abruptly between $\beta=6.0$ and
$\beta=5.9$. The $\chi^2/dof$ value is acceptable only for an
$\alpha$-cut where $\alpha=0.5$.

The Berlin data allows only to compare $\beta=6.0$ and $\beta=5.8$ to
$\beta=6.2$ (which sets the reference scale). The ratios of the lattice 
spacings are compatible with the results for the Osaka data.     
The ratio of the renormalization constants is still compatible with
unity for $\beta=6.0$, if compared to $\beta=6.2$, but drops 
between $\beta=6.0$ and $\beta=5.8$ similar to the Osaka data.
The $\chi^2/dof$ is unacceptably large.

\begin{figure}
\vspace{0.2cm}
\centering
  \includegraphics[width=0.8\linewidth]{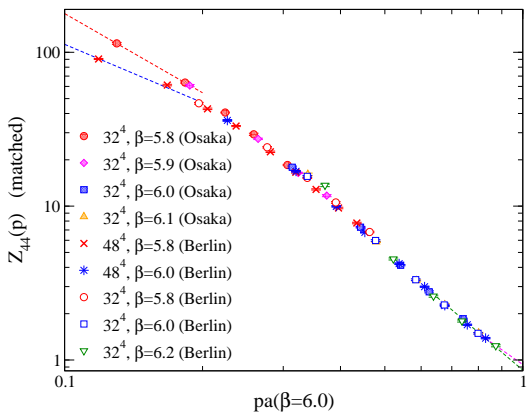}
  \caption{The dressing function of the instantaneous time-time gluon
    propagator after matching Berlin and Osaka data separately on 
    $32^4$ and $48^4$ lattices for five different $\beta$ values.
    Fit parameters underlying the matching are listed in 
    Tabs.~\ref{tab:table5} and \ref{tab:table6}. Data has been  
    cylinder and cone cut and results are shown for a fixed 
    $\alpha$-cut with $|p_i a| \le 0.5$. The two fits for the 
    momentum dependence in the IR are described in \Sec{sec:fits}.}
\label{fig:fig9}
\end{figure}

In~\Fig{fig:fig9} we show our final result for the instantaneous
time-time gluon propagator having matched and combined all the Osaka and
Berlin data. We have now used a unique scale set by $a(\beta=6.0)$ to
give all momenta in physical units. The quality of the fits in this
case is worse compared with the fits of the transverse gluon
propagator. Nevertheless, the scaling behavior does look quite
reasonable and there is an improvement compared to the results
presented in \Tab{tab:table4}. The reason is probably that we have
moved closer to the continuum limit by including data from $\beta=6.1$
and $\beta=6.2$. 
In the infrared region the data points obtained in Berlin and Osaka 
split. We interprete this as a consequence of the use of different gauge 
fixing techniques. The more efficient simulated annealing method weakens
the singular behavior as seen also for the ghost propagator 
\cite{Voigt:2008rr}.

\begin{figure*}
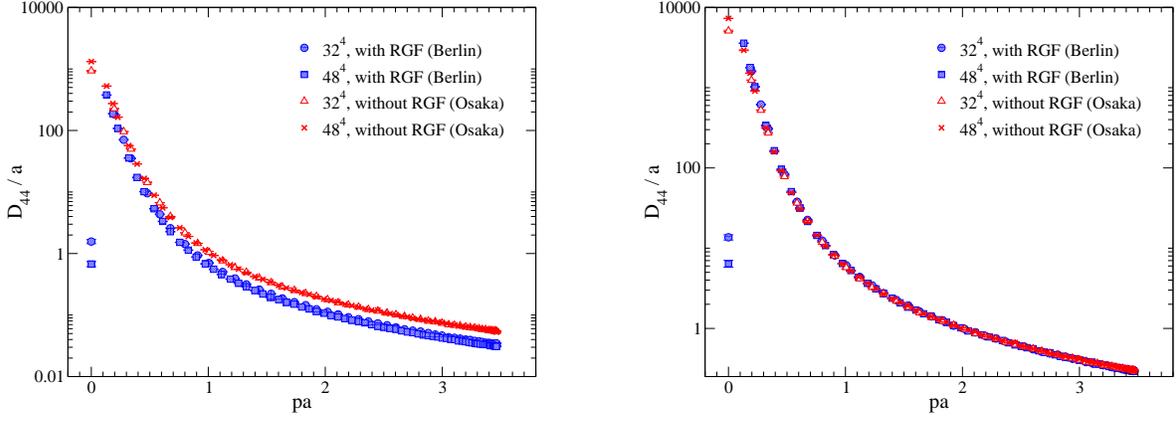

 \centering
 \mbox{\includegraphics[height=5.5cm]{D44_comparison_rawdata}
\qquad\qquad
 \includegraphics[height=5.5cm]{D44_comparison_normalized}}
   \caption{ Left: comparison between the unrenormalized instantaneous 
     time-time gluon propagator $D_{44}$ with and without the residual gauge 
     fixing.  Right: the same but formally normalized at $pa=2$.
     The cylinder cut has applied. Berlin and Osaka data for $\beta=6.0$ 
     and lattice sizes $32^4$ and $48^4$ are shown together.}
\label{fig:fig10}
\end{figure*}

In passing, we revisit the question whether there
is a difference between the instantaneous time-time gluon propagator
if one is applying the residual gauge-fixing (Berlin data) or not
(Osaka data). \Fig{fig:fig10} shows the results for $\beta=6.0$. The
propagator at momentum $pa \ne 0$ seems not to depend on the volume,
but on the procedure (cf.\ the left panel).  The latter is
understandable if one looks back at \Fig{fig:fig1} and there at the
difference for the time-like links. It is remarkable that the
difference between the two cases can be eliminated by a uniform
multiplicative rescaling. This is accomplished by normalizing the
propagator $D_{44}/a$ at $pa = 2.0$ to 1.0 (cf.\ the right panel).
The residual gauge fixing has only an impact on the value of the
propagator at zero momentum, $D_{44}(ap=0)/a$.  With residual gauge
fixing this value is obviously smaller as expected.

\section{Matching the ghost propagator}
\label{sec:ghost}

Finally we apply the matching procedure to the ghost propagator. As we
have seen in \Fig{fig:fig4}, Necco-Sommer scaling is only weakly
violated. Therefore, we expect a matching result that closely follows
this behavior.  
\begin{figure}[h!]
\vspace{0.2cm}
  \centering
   \includegraphics[width=0.8\linewidth]{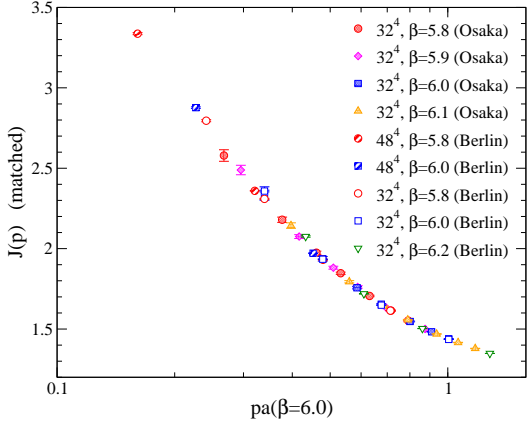}
   \caption{The ghost dressing function after matching data from
     $32^4$ and $48^4$ lattices including data from both Osaka and
     Berlin at five different $\beta$ values. For the fits see
     \Tab{tab:table7} and \Tab{tab:table8}. The result is shown for a
     fixed $\alpha$-cut: $\abs{p_i a} \le 0.5$. Both the cylinder and the
     cone cut are applied as usual.} 
\label{fig:fig11}
\end{figure}

The fitting results (respectively for the Osaka and Berlin data) are
presented in Tabs.~\ref{tab:table7} and \ref{tab:table8}, there
relative to the highest $\beta_{\rm max}=6.1$ and $\beta_{\rm
  max}=6.2$ in both cases. For no $\alpha$-cut ($\alpha=2.0$) the
ratios of lattice spacing $a(\beta)/a(\beta_{\rm max})$ reproduce
almost perfectly the corresponding ratios according to
\Eq{eq:necco-sommer-interpolation}. Nevertheless, we notice that a
smaller $\alpha$ leads to some deviation from that scaling, in
particular at the lowest $\beta$. Including the results of the
separate fits, the Osaka and Berlin data are afterwards combined in
\Fig{fig:fig11} showing there the result only for the most  
restrictive $\alpha$-cut ($\alpha=0.5$). 

Matching the Osaka data yields an overall very good $\chi^2/dof$. The
ratios of the renormalization constants are all compatible with unity,
and the ratios of lattice spacings rise monotonously upon decreasing
$\beta$. When no $\alpha$-cut is applied the fitted $R_a$ rise in
accordance with \Eq{eq:necco-sommer-interpolation}, while restricted
$\alpha$-cuts lead to $R_a$'s which grow slightly slower.
Matching the Berlin data results in the same tendencies, but the
$\chi^2/dof$ turned out to be very large.
Probably due to the fact that the Berlin ghost data are averaged
over all time slices and the Osaka data are not, the errors of the Berlin
data are smaller by an order of magnitude. 

\section{The scaling behavior of the propagators}
\label{sec:scaling}

\begin{figure*}
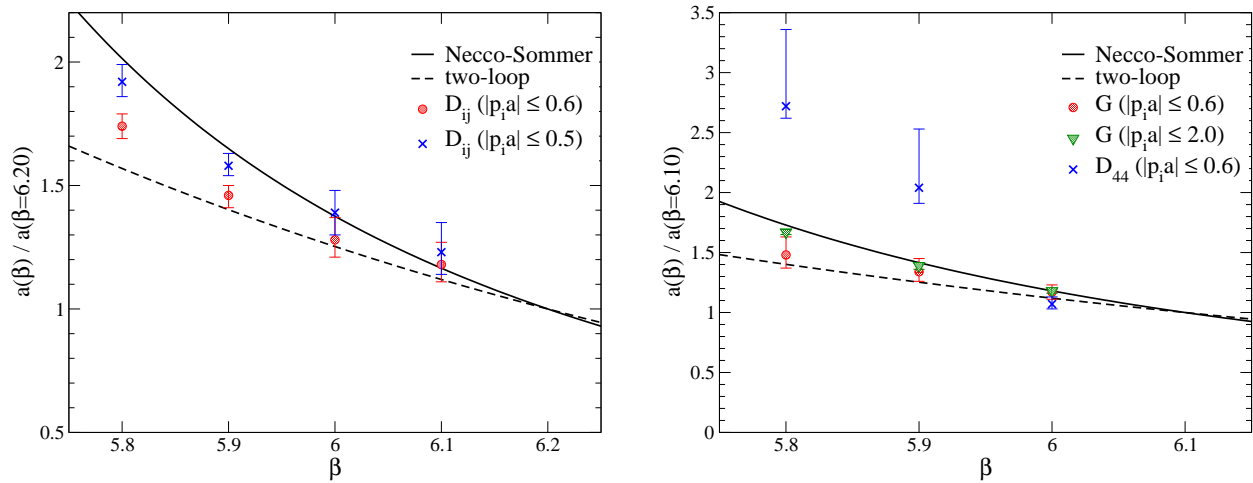

\vspace{0.2cm}
  \centering
  \mbox{\includegraphics[width=0.44\linewidth]{ScalingRelation_580to620}
    \qquad
    \includegraphics[width=0.44\linewidth]{ScalingRelation_580to610}}
  \caption{The scaling relation as it emerges from matching the data
    obtained in simulations with the Wilson action. Left: for the
    transversal gluon propagator with two $\alpha$-cuts for
    comparison. Right: for the ghost propagator with no $\alpha$-cut
    ($\alpha=2.0$) and an $\alpha$-cut with $\alpha=0.6$ for
    comparison, and for the time-time gluon propagator with an
    $\alpha$-cut with $\alpha=0.6$, a rather conservative momentum
    cut.} 
\label{fig:fig12}
\end{figure*}

At this point we can check now if our individual results on $a(\beta)$
reproduce a unique running lattice scale.  
We had started with Necco-Sommer scaling, but abandoned this, fully
relying on the matching procedure to produce the ``correct'' lattice
scale function. Let us remind that the lattice scales as found
may deviate from asymptotic scaling (which of course is strictly valid
only for $\beta \to \infty$) and also from that derived for other
observables (e.g., Necco-Sommer scaling derived for the static
quark-antiquark force in pure $SU(3)$ gauge theory). 

With \Fig{fig:fig12} we summarize our results on the fitted
scaling behavior of the lattice spacing in terms of the finest lattice
spacing available in our study. For all the propagators considered we
have plotted the ratios of the lattice spacings relative to the finest
lattice (at $\beta=6.2$ for the transverse gluon propagator in the
left panel, and at $\beta=6.1$ for $D_{44}$ and the ghost propagators
in the right panel, respectively) as a function of $\beta$ as found
through the matching procedure with two choices of the
$\alpha$-cut. For the transversal gluon propagator the data points
(corresponding to both choices of $\alpha$-cuts) fit very well between
the curves corresponding to two-loop running of the lattice spacing,
\begin{equation}
  a = \frac{1}{\Lambda_{\textrm{Lat}}}
  \left(\frac{8\pi^2}{11N_c}\beta\right)^{\frac{51}{121}}
  \exp\left( -\frac{4\pi^2}{11N_c}\beta\right) \, .
\end{equation}
and to the relation (\Eq{eq:necco-sommer-interpolation}). The same is
observed for the ghost propagator. In this case, also applying no cut
($\alpha=2.0$) results in a reasonable result. 

On the other hand, the $a(\beta)$ dependence as found from the
matching procedure of the instantaneous time-time gluon propagator is
not only much stronger than in \Eq{eq:necco-sommer-interpolation}, but
also inconsistent with the scaling law for the other propagators. We
cannot say to which extent such a faster running is beyond some general
bound and unfortunately have to conclude that the problem of the bad
scaling behavior for the instantaneous time-time gluon propagator
remains unsolved yet. 

\section{Fitting the behavior at large and small momentum}
\label{sec:fits}
Having successfully merged data for the propagators from simulations
at different $\beta$ values, one may try to fit their ultraviolet (UV)
behavior and partly also to extract some infrared (IR) exponents. 

For the transverse gluon propagator we try a power-law ansatz 
\begin{equation}
 D^{\rm tr}(\abs{\vec{p}})^{\mathrm{UV}} =
 \frac{1}{|\vec{p}|} \left( \frac{c_{\rm tr}}{|\vec{p}|} \right)^{\eta_{\rm tr}}, 
\end{equation}
to describe the behavior at large momenta. For the UV fitting we use
the data points above some minimal momentum
($\abs{\vec{p}}_{\mathrm{min}}$ in units of $a^{-1}(\beta=6.2)$) and 
investigate the dependence of the anomalous dimension $\eta_{\rm tr}$ 
on the fitting range and the $\alpha$-cut. Fit results are collected   
in~\Tab{tab:table9} and our best fits (with
$\alpha=0.5$) give $\eta_{\rm tr} =0.40(2)$. Qualitatively, the behavior 
we find is similar to the UV fit given in Ref.~\cite{Langfeld:2004qs},
though there [for $SU(2)$] a somewhat bigger exponent $\eta_{\rm tr}=0.5(1)$ 
was found.  

For the longitudinal gluon dressing function $Z_{44}$ we try power law 
ansatzes both in the UV and in the IR regions
\begin{equation}
Z_{44}(|\vec{p}|)^{\mathrm{UV}} = \left( \frac{c_{44}}{|\vec{p}|} \right)^{\eta_{44}}\,, ~~~
Z_{44}(|\vec{p}|)^{\mathrm{IR}} = \left( \frac{d_{44}}{|\vec{p}|} \right)^{\kappa_{44}}\,.
\end{equation}
The results are collected in Tabs. \ref{tab:table10} and \ref{tab:table11}, 
respectively. This dressing function was not studied in Ref.~\cite{Langfeld:2004qs}.

For the ghost dressing function, analogous to Ref.~\cite{Langfeld:2004qs},
we adopt a logarithmic ansatz in the UV region
\begin{equation}
   J(|\vec{p}|)^{\mathrm{UV}} = 
    \frac{c_{\rm gh}}{\ln(|\vec{p}|/\Lambda_{\textrm{Coul}})^{\gamma}}\,,
\end{equation}
and a power-law ansatz for the IR behavior
\begin{equation}
  J(|\vec{p}|)^{\mathrm{IR}} = \left( \frac{d_{\rm gh}}{|\vec{p}|} \right)^{\kappa_{\rm gh}}\,.
\end{equation}

Fits results for either momentum region are given in Tabs.~\ref{tab:table12} 
and \ref{tab:table13}, respectively. The UV fits scatter with $\alpha$ (the
additional momentum cut), though, the most stable results are obtained for no 
$\alpha$-cut ($\alpha=2.0$). With a suitably restricted fit interval fits are 
stable and give $\Lambda_{\rm Coul}a(\beta=6.0)=0.275(20)$ or 
$\Lambda_{\rm Coul} r_0=1.37(10)$  and $\gamma=0.33(1)$. For $SU(2)$ this 
exponent was found to be $\gamma=0.26(2)$ \cite{Langfeld:2004qs}. 

The IR fits are quite stable and give $\kappa_{\rm gh}=0.435(6)$ without
applying $\alpha$-cuts, even though we admit that the $\chi^2/dof$ values 
are rather large. In Ref.~\cite{Langfeld:2004qs} a value 
$\kappa_{\rm gh}=0.49(1)$ was found (corresponding to $2\kappa$ there). 

\section{Conclusions}
\label{sec:conclusion}

We have investigated the momentum dependence of the instantaneous
ghost and gluon propagators of pure $SU(3)$ lattice Coulomb gauge 
theory. Our study represents a joint analysis of data from lattice
simulations independently performed at Berlin and Osaka for the Wilson 
gauge action in the range $\beta=5.8,\ldots,6.2$.  

For these values of $\beta$, we find apparent scaling violations
for both the spatially transversal and the time-time gluon propagator,
while for the ghost propagator such violations are surprisingly
mild. Our inspection of the gluon propagator data shows that the
violations there are basically due to data that survives a cylinder
cut but involves momentum components close to the upper end of the
Brillouin zone. Consequently, if additionally an $\alpha$-cut like 
$\abs{p_ia} \le 0.5$ is applied to the data, scaling violations are
under much better control. The price to pay are strong restrictions of
allowed momenta which, in our opinion, should not only satisfy the
cylinder and cone cuts but also $\abs{p_ia} \le 0.5$
($\alpha$-cut). This is the first result of our paper.

Second, we find that the scaling violations can be sufficiently
reduced if, in addition to the aformentioned cuts, a matching
procedure (see \App{app:matching}) is used to merge data. 
That is, instead of imposing one particular $a(\beta)$ dependence
(e.g., that of Ref.~\cite{Necco:2001xg}) and normalizing the data for
the different lattice cutoffs such that they coincide at a particular 
reference scale, both the $a(\beta)$ dependence and the relative
normalization factors are determined through an optimization method
that seeks the best overlap of data. It turns out, that the matching  
procedure applied to either the transversal gluon or the ghost
propagator provides us with a $a(\beta)$ dependence only slightly
different from what is known from~\cite{Necco:2001xg}, somewhere in
between Necco-Sommer scaling and asymptotic two-loop scaling. Note
that the matching procedure would allow us to fix the lattice 
spacing if we were to simulate also beyond the interval $5.7 \le \beta
\le 6.92$ covered by the Necco-Sommer analysis. 

Generally we can say that the matching analysis results in ratios
of the renormalization constants closer to unity at $\beta \ge 6.0$.
Future lattice studies of gluon and ghost propagators should be
performed in that region. The fact that -- except for the ghost
propagator -- the matching performs better the more restrictive
$\alpha$-cuts are applied shows that the momenta with components
close to the upper end of the Brillouin zone are far from 
the continuum limit. This might signal a more general effect, namely
that observables closer to the infrared region have better scaling
properties. 

Unfortunately, we could not correct the scaling violations for the
instantaneous time-time gluon propagator. For this, these violations
are so strong that the $a(\beta)$ dependence as found through
the matching is far from what we find for the other propagators. In
fact, $a(\beta)$ in this case is found running too fast. Moreover, for
$\beta \le 5.9$ the ratio of renormalization constants
drops compared to the behavior at $\beta \ge 6.0$ such that the
assumptions and results of the matching analysis for the $D_{44}$
propagator must be considered with caution.

We mention that for the $SU(2)$ transversal gluon propagator it has
been argued~\cite{Burgio:2008jr} that the correct instantaneous
propagator can be reconstructed only from the full 4-dimensional
space-time propagator.  There, a residual gauge-fixing was applied
that enforces 
$A_4=\mathrm{const}$. Therefore, it needs to be
scrutinized whether the scaling violations, that we have seen here for
the transversal gluon propagator, are really due to the alleged
(multiplicative) non-renormalizability of the Coulomb
gauge~\cite{Watson:2007mz} when residual gauge-fixing is applied or
not. Our results for the transversal propagator suggest a more mundane
resolution: exclude too large momenta from the analysis and allow for
an independently determined running lattice spacing, then data within
a very restricted range of momenta (in units $a$) can be successfully
merged and gives a $a(\beta)$ dependence that agrees with what is
known form the literature.

We stress again that our result for the time-time gluon propagator
is non-acceptable. The $a(\beta)$ dependence as found for this is far
from the running scale for the other propagators. In the light of this, the
argument of non-renormalizability might still be valid for the 
$A_4$ component of the gluon field.

When applying fits to the data at either low or large momenta
(though restricted by quite stringent bounds) we obtain qualitatively
similar UV and IR fits as reported for the $SU(2)$ theory
in~\cite{Langfeld:2004qs}.

\section*{Acknowledgements}
Simulations were performed on a SX-8R (NEC) vector-parallel computer
at the RCNP of Osaka University and on a IBM p690 system at HLRN,
Berlin and Hannover, Germany. We appreciate the warm hospitality and
support of the RCNP and HLRN administrators. 
We thank Hinnerk Stueben for contributing parts of the code used at 
HLRN and help for performing simulations there.
This work is partly supported by Grants-in-Aid for Scientific Research 
from Monbu-Kagaku-sho (No. 17340080 and 20340055).  Y.~N. is supported by
Grant-in-Aid for JSPS Fellows from the Ministry of Education, Culture,
Sports, Science and Technology of Japan, and A.~S. by the Australian
Research Council.  The work of E.-M.~I. was supported by DFG through
the Forschergruppe FOR 465 (Mu932/2). He is grateful to the
Karl-Franzens-Universit\"at Graz for the hospitality while
this paper was being completed.  E.-M.~I., M.~M.-P. and Y.~N. gratefully
acknowledge useful discussions with G. Burgio and P. Watson.



\begin{appendix}

\section{Matching procedure}
\label{app:matching}

\begin{figure*}
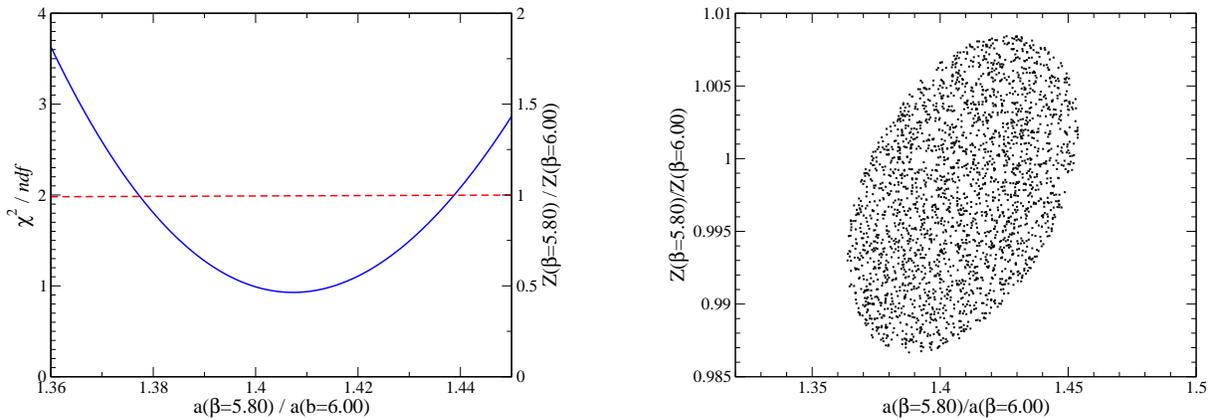

\vspace{0.2cm}
   \mbox{\includegraphics[height=5.5cm]{RaRz_PhysV_050}
     \qquad\qquad
     \includegraphics[height=5.5cm]{RaRz_PhysV_050_confident}}
   \caption{Details of matching $Z^{\mathrm{tr}}$ measured on a
     $L^4=32^4$, $\beta=5.8$ lattice with the data obtained on a
     $L^4=48^4$, $\beta=6.0$ lattice. An $\alpha$-cut with $\abs{p_i
       a} \le 0.5$ was applied before matching. Left:
     $\chi^2/dof$ as a function of the ratio of lattice  
     spacings $R_a = a_c(5.8) / a_f(6.0)$. 
     Right: the 68.3\% confidence region spanned by ratios of lattice
     spacing $R_a = a_c(5.8) / a_f(6.0)$ and
     renormalization constants $R_Z = Z(5.8) / Z(6.0)$ 
     determined through the matching procedure.}
\label{fig:fig13}
\end{figure*}

In this appendix we describe the matching procedure of
\cite{Leinweber:1998uu} applied to Coulomb gauge. The procedure does
not rely on any given lattice scale dependence $a(\beta)$ but allows
us to extract this for each propagator individually.

Under the assumption that the fixed-time gluon and ghost propagators
in Coulomb gauge can be renormalized multiplicatively (see
\Sec{sec:propagators} and Ref.~\cite{Watson:2007vc}), we aim at an
optimal overlap of bare propagator data from a coarse lattice (with
unknown lattice spacing $a_{c}$) and a fine lattice (with a lattice
spacing $a_{f}$ that might be known). Using the fact that the bare and 
dimensionless lattice propagator $D^L$ is a function of the product of
the three-momentum $p$ with the lattice spacing $a$ only (the
dependence on $\beta$ is of course kept in mind), and assuming that
multiplicative renormalization is valid, the bare propagators on the
fine and coarse lattice are related by 
\begin{equation}
  a_f D^L_f(p a_f) = R_Z(a_f / a_c)~\cdot~a_c D^L_c(p a_c) 
\end{equation}
The renormalization factor $R_Z$ only depends on the ratio of the lattice
cutoffs $R_a = a_f / a_c$. Taking the logarithm gives
\begin{equation}
  \ln D^L_f(pa_f) = \ln D^L_c(pa_c) - \ln R_a + \ln R_Z \, .
\end{equation}
Expressing the momentum on the coarse lattice in terms of the
momentum on the fine lattice by 
\begin{equation}
  a_c = \frac{a_f}{R_a} \quad\Longleftrightarrow\quad \ln(pa_c) =
  \ln(pa_f) - \ln R_a \, , 
\end{equation}
we arrive at
\begin{align}
  \ln D^L_f[\ln(pa_f)]
  &= \ln D^L_c [\ln(pa_f) - \ln R_a ] - \ln R_a + \ln R_Z \nonumber \\
  &= \ln D^L_c [\ln(pa_f) + \Delta_a ] + \Delta_Z \, ,
\label{eq:condition}
\end{align}
where $R_a = e^{-\Delta_a}$ and $R_Z = e^{-\Delta_a +\Delta_Z}$.

Notice that $\Delta_a$ and $\Delta_Z$ are positive.  We find the
values for $R_a$ and $R_Z$ from a fitting procedure as follows.

Suppose that we have one data set $\{x=pa_f,D^L_{f},\sigma_{f}\}_{i}$
with $i = 1,\ldots, n_f$ for the fine lattice and one data set
$\{y=pa_c,D^L_{c},\sigma_{c}\}_{j}$ with $j = 1,\ldots, n_c$ for the
coarse lattice with $\sigma$ denoting the statistical error of the
propagator $D^L$, and $n_f$ resp. $n_c$ denoting the number of data
points for the propagator on the fine and coarse lattice,
respectively.  Then, we use a $\chi^{2}$ fit to optimally match both
data sets, i.e., to find the optimal overlap of the bare lattice
propagator from the fine and the coarse lattice. 
To be specific, we minimize
\begin{widetext}
\begin{equation}
\chi^2 = \sum_{i=1}^{n_f}
\left(\frac{D^L_{\textrm{f}}(x_i)
 - \frac{R_Z}{R_a}D^{L~\textrm{int}}_{\textrm{c}}\left(\frac{x_i}{R_a}\right)}
{\sigma_{\textrm{f},i}}\right)^2
+ \sum_{j=1}^{n_c}
\left(\frac{D^L_{\textrm{c}}(y_j)
 - \frac{R_a}{R_Z}D^{L~\textrm{int}}_{\textrm{f}}\left(y_j R_a\right)}
{\sigma_{\textrm{c},j}}\right)^2 \, .
\end{equation}
\end{widetext}
In the first term $D^L_{\textrm{f}}$ is represented by the measured
values at the momenta $p_i$ (expressed as function of $x_i=p_i a_f$)
and the corresponding error $\sigma_{\textrm{f},i}$, while
$D_{\textrm{c}}^{L~\textrm{int}}$ is evaluated at these momenta by a
cubic spline interpolation of the data for $D_{\textrm{c}}^L$ .  In
the second term, the r\^ole of $D_{\textrm{f}}^L$ and
$D_{\textrm{c}_L}$ is interchanged with respect to genuine data
($y_j=p_j a_c$ in $D_{\textrm{c}_L}$) and interpolation of
$D_{\textrm{f}}^L$.  With this definition of $\chi^2$ the matching is
done as follows:
\begin{enumerate}
\item Vary $\Delta_a$ over an interval $(0,1]$ with step size $0.001$
  and determine the optimal $\Delta_Z$ giving the lowest $\chi^2/dof$ for
  each value of $\Delta_a$ considered,
\item Identify the best overall combination of $\Delta_a$ and
  $\Delta_Z$ by searching for the global minimum of $\chi^2/dof$. 
\end{enumerate} 
This provides us with the optimal choice of $R_a$ and $R_Z$.  The
error of $R_a$ and $R_Z$ are given by the 68.3\% confidence region,
i.e., the region of fit parameters $R_a$ and $R_Z$ with $\chi^2/dof <
\chi^2_{\mathrm{min}}/dof + 1$. An illustration of this is given in
\Fig{fig:fig13} for matching the instantaneous transversal gluon
propagator measured on a $L^4=32^4$, $\beta=5.8$ lattice with data
obtained on a $L^4=48^4$, $\beta=6.0$ lattice
(cf.\ \Sec{sec:transversal_gluon}, \Fig{fig:fig6}
and~\Tab{tab:table2}).

Note that applying this procedure to several combinations of fine and
coarse lattices provides us with an optimal scaling relation
$a=a(\beta)$ for each propagator. \\~\\

\section{Tables}
\label{app:tables}

In this appendix we present an overview of the data sets produced
in Osaka and Berlin, the results of all matching fits according
to Secs. \ref{sec:transversal_gluon}, \ref{sec:time_time_gluon},
\ref{sec:ghost} and \App{app:matching} as well as of the fits
in the infrared (IR) and ultraviolet (UV) limits as described in 
\Sec{sec:fits}.
\begin{table}[b]
  \caption{Lattice parameters used in this study. Configurations were
    generated at RCNP Osaka and HU Berlin.}
  \label{tab:table1}
  \begin{tabular}{c@{\quad}c@{\quad}c@{\quad}c@{\quad}c@{\quad}r@{\quad}r}
    \hline\hline\\*[-2.0ex]
    $L^4$ &   $\beta$   & $a^{-1}$ [GeV] & $a$ [fm] &$V$[fm$^4$]& \#conf
    & group \\*[1.0ex]
    \hline\\*[-2ex]
 $12^4$ & 5.8 & 1.446 & 0.1364 & 1.64$^4$  & 100 & Berlin  \\
 $16^4$ &  :  &   :   & :      & 2.18$^4$  & 40  & Berlin  \\
 $18^4$ &  :  &   :   & :      & 2.46$^4$  & 80  & Osaka   \\
 $24^4$ &  :  &   :   & :      & 3.27$^4$  & 40  & Osaka   \\
 $24^4$ &  :  &   :   & :      & 3.27$^4$  & 30  & Berlin  \\
 $32^4$ &  :  &   :   & :      & 4.36$^4$  & 20  & Osaka   \\
 $32^4$ &  :  &   :   & :      & 4.36$^4$  & 30  & Berlin  \\
 $48^4$ &  :  &   :   & :      & 6.55$^4$  & 20  & Berlin  \\*[1ex]

 $18^4$	& 5.9 & 1.767 & 0.1116 & 2.09$^4$  & 80  & Osaka   \\
 $24^4$	&  :  &   :   &   :    & 2.78$^4$  & 40  & Osaka   \\
 $32^4$	&  :  &   :   &   :    & 3.71$^4$  & 20  & Osaka   \\*[1ex]

 $12^4$	& 6.0 & 2.118 & 0.0932 & 1.12$^4$  & 100 & Berlin  \\     
 $16^4$ &  :  &   :   &   :    & 1.49$^4$  &  60 & Berlin  \\
 $18^4$ &  :  &   :   &   :    & 1.68$^4$  &  80 & Osaka   \\
 $24^4$ &  :  &   :   &   :    & 2.24$^4$  &  40 & Osaka   \\
 $24^4$ &  :  &   :   &   :    & 2.24$^4$  &  40 & Berlin  \\
 $32^4$ &  :  &   :   &   :    & 2.98$^4$  &  20 & Osaka   \\
 $32^4$ &  :  &   :   &   :    & 2.98$^4$  &  30 & Berlin  \\
 $48^4$ &  :  &   :   &   :    & 4.48$^4$  &  20 & Berlin  \\*[1ex]

 $18^4$ & 6.1 & 2.501 & 0.0788 & 1.42$^4$  &  80 & Osaka   \\
 $24^4$ &  :  &   :   &   :    & 1.89$^4$  &  40 & Osaka   \\
 $32^4$ &  :  &   :   &   :    & 2.52$^4$  &  20 & Osaka   \\*[1ex]

 $12^4$ & 6.2 & 2.914 & 0.0677 & 0.81$^4$  &  100 & Berlin \\ 
 $16^4$ &  :  &   :   &   :    & 1.08$^4$  &  40  & Berlin \\
 $24^4$ &  :  &   :   &   :    & 1.62$^4$  &  30  & Berlin \\
 $32^4$ &  :  &   :   &   :    & 2.17$^4$  &  20  & Berlin \\*[0.5ex]
\hline\hline
\end{tabular} 
\end{table}
\begin{table}[b]
\centering
\caption{Fit parameters obtained upon matching $D^{\mathrm{tr}}$ data
  from a $(L^4,\beta)=(32^4,5.8)$ and a $(48^4,6.0)$ lattice for two
  different $\alpha$-cuts (see \Fig{fig:fig6}). For comparison we also
  show the lattice spacing ratios according to
  \Eq{eq:necco-sommer-interpolation}.}  
\begin{tabular}{c@{\qquad}c@{\quad}c@{\quad}c@{\quad}c}
\hline\hline\\*[-2.0ex]
 $\abs{p_i a} \le \alpha$ & $\frac{a(5.8)}{a(6.0)}$ & 
 $\frac{a^{\mathrm{NS}}(5.8)}{a^{\mathrm{NS}}(6.0)}$ & 
 $R_Z\left(\frac{a(5.8)}{a(6.0)}\right)$ & $\chi^2/dof$ \\*[1.0ex]
\hline\\*[-2ex]
    $\alpha=0.6$ & $1.37^{+3}_{-4}$ & $1.46$ & $0.989^{+10}_{-10}$ & $3.28$ \\
    $\alpha=0.5$ & $1.41^{+4}_{-5}$ & $1.46$ & $0.998^{+11}_{-11}$ & $0.92$ \\*[0.5ex]
\hline\hline
\end{tabular}
\label{tab:table2}
\end{table}
\begin{table}
\centering
\caption{Matching the transversal gluon propagator for five $\beta$
  values (see Fig.~\ref{fig:fig7}): shown are the ratios of lattice
  spacings relative to the finest one obtained either by the matching
  procedure or according to \Eq{eq:necco-sommer-interpolation}; the
  ratios of the renormalization constants and the corresponding
  $\chi^2/dof$ of the fit that accomplishes the matching, for four
  choices of the $\alpha$-cut. The lattice size is $32^4$.}
\begin{tabular}{c@{\qquad}c@{\qquad}c@{\quad}c@{\quad}c@{\quad}c}
\hline\hline\\*[-2.0ex]
$|p_i a| \le \alpha$ & $\beta$ & $\frac{a(\beta)}{a(6.2)}$ & 
$\frac{a^{\mathrm{NS}}(\beta)}{a^{\mathrm{NS}}(6.2)}$ & 
$R_Z(\frac{a(\beta)}{a(6.2)})$ & $\chi^2/dof$ \\[1.0ex] 
\hline\\*[-2ex]
\multirow{4}{*}{$\alpha=1.0$}
& $5.8$ & $1.51^{+4}_{-4}$   & $2.01$ & $0.982^{+12}_{-13}$  & $5.56$ \\
& $5.9$ & $1.31^{+5}_{-5}$   & $1.65$ & $0.990^{+19}_{-18}$  & $3.83$ \\
& $6.0$ & $1.19^{+7}_{-7}$   & $1.38$ & $0.993^{+34}_{-26}$  & $2.39$ \\
& $6.1$ & $1.12^{+10}_{-20}$ & $1.17$ & $0.984^{+108}_{-36}$ & $6.30$ \\*[0.5ex]

\multirow{4}{*}{$\alpha=0.8$}
& $5.8$ & $1.68^{+4}_{-6}$   & $2.01$ & $0.961^{+12}_{-10}$  & $2.16$ \\
& $5.9$ & $1.43^{+5}_{-6}$   & $1.65$ & $0.970^{+16}_{-15}$  & $2.98$ \\
& $6.0$ & $1.27^{+7}_{-9}$   & $1.38$ & $0.973^{+35}_{-23}$  & $2.61$ \\
& $6.1$ & $1.18^{+8}_{-13}$  & $1.17$ & $0.968^{+52}_{-25}$  & $5.26$ \\*[0.5ex]

\multirow{4}{*}{$\alpha=0.6$}
& $5.8$ & $1.74^{+5}_{-5}$   & $2.01$ & $0.974^{+9}_{-9}$    & $1.41$ \\
& $5.9$ & $1.46^{+4}_{-5}$   & $1.65$ & $0.980^{+12}_{-11}$  & $1.70$ \\
& $6.0$ & $1.28^{+9}_{-7}$   & $1.38$ & $0.983^{+23}_{-22}$  & $2.97$ \\
& $6.1$ & $1.18^{+9}_{-7}$   & $1.17$ & $0.976^{+27}_{-27}$  & $1.87$ \\*[0.5ex]

\multirow{4}{*}{$\alpha=0.5$}
& $5.8$ & $1.92^{+7}_{-6}$   & $2.01$ & $0.963^{+8}_{-8}$    & $0.585$ \\
& $5.9$ & $1.58^{+5}_{-4}$   & $1.65$ & $0.967^{+9}_{-9}$    & $2.49$ \\
& $6.0$ & $1.39^{+9}_{-9}$   & $1.38$ & $0.970^{+19}_{-19}$  & $1.96$ \\
& $6.1$ & $1.23^{+12}_{-9}$  & $1.17$ & $0.968^{+27}_{-26}$  & $1.59$ \\*[0.5ex]  
\hline\hline
\end{tabular}
\label{tab:table3}
\end{table}
\begin{table}
\caption{Fit parameters obtained upon matching the instantaneous
  time-time gluon propagator on a $L^4=32^4$, $\beta=5.8$ and a
  $L^4=48^4$, $\beta=6.0$ lattice with two different $\alpha$-cuts (see
  \Fig{fig:fig8}).  For comparison we show also the lattice spacing
  ratio predicted by the Necco-Sommer scaling relation.}
\centering
\begin{tabular}{c@{\qquad}c@{\quad}c@{\quad}c@{\quad}c}
\hline\hline\\*[-2.0ex]
$|p_i a| \le \alpha$ & $\frac{a(5.8)}{a(6.0)}$ & 
$\frac{a^{\mathrm{NS}}(5.8)}{a^{\mathrm{NS}}(6.0)}$ &
$R_Z\left(\frac{a(5.8)}{a(6.0)}\right)$ & $\chi^2/dof$ \\*[1.0ex]
\hline\\*[-2ex]
$\alpha = 0.6$ & $1.96^{+17}_{-9}$  & $1.46$ & $0.476^{+48}_{-55}$ & $4.50$ \\
$\alpha = 0.5$ & $1.87^{+12}_{-10}$ & $1.46$ & $0.502^{+42}_{-40}$ & $0.503$ \\*[0.5ex]
\hline\hline
\end{tabular}
\label{tab:table4}
\end{table}
\begin{table}
\caption{Matching the time-time gluon propagator for four $\beta$
  values on $32^4$ lattices (Osaka data): shown are the ratios of
  lattice spacings obtained by the matching relative to the finest one,
  for comparison also the ratios predicted by Necco-Sommer scaling, the
  ratios of the renormalization constants and the corresponding
  $\chi^2/dof$ of the fit that accomplishes the matching, for two
  choices of the $\alpha$-cut. }
\centering
\begin{tabular}{c@{\qquad}c@{\quad}c@{\quad}c@{\quad}c@{\quad}c}
\hline\hline\\*[-2.0ex]
$|p_i a| \le \alpha$ & $\beta$ & $\frac{a(\beta)}{a(6.1)}$ & 
$\frac{a^{\mathrm{NS}}(\beta)}{a^{\mathrm{NS}}(6.1)}$ &
$R_Z\left(\frac{a(\beta)}{a(6.1)}\right)$ & $\chi^2/dof$ \\*[1.0ex]
\hline\\*[-2ex]
\multirow{3}{*}{$\alpha=0.6$}
  & $5.8$ & $2.72^{+64}_{-10}$ & $1.73$ & $0.327^{+19}_{-84}$  & $4.94$ \\
  & $5.9$ & $2.04^{+49}_{-13}$ & $1.42$ & $0.429^{+44}_{-128}$ & $13.9$ \\
  & $6.0$ & $1.07^{+6}_{-4}$   & $1.18$ & $1.04^{+7}_{-7}$     & $8.70$ \\*[1.0ex]

\multirow{3}{*}{$\alpha=0.5$}
  & $5.8$ & $2.60^{+18}_{-26}$ & $1.73$ & $0.350^{+53}_{-32}$  & $1.45$ \\
  & $5.9$ & $1.81^{+5}_{-18}$  & $1.42$ & $0.509^{+96}_{-29}$  & $1.03$ \\
  & $6.0$ & $1.08^{+8}_{-8}$   & $1.18$ & $1.03^{+15}_{-12}$   & $6.19$ \\*[0.5ex]
\hline\hline
\end{tabular}
\label{tab:table5}
\end{table}
\begin{table}
  \caption{Matching the time-time gluon propagator for three $\beta$
    values on $32^4$ lattices (Berlin data): shown are the ratios of lattice 
    spacings obtained by the matching relative to the finest one, 
    for comparison also the ratios predicted by Necco-Sommer scaling, 
    the ratios of the renormalization constants 
    and the corresponding $\chi^2/dof$ of the fit 
    that accomplishes the matching, for two choices of the $\alpha$-cut. }
\centering
\begin{tabular}{c@{\quad}c@{\quad}c@{\quad}c@{\quad}c@{\quad}c}
\hline\hline\\*[-2.2ex]
$|p_i a| \le \alpha$ & $\beta$ & $\frac{a(\beta)}{a(6.2)}$ & 
$\frac{a^{\mathrm{NS}}(\beta)}{a^{\mathrm{NS}}(6.2)}$ &
$R_Z\left(\frac{a(\beta)}{a(6.2)}\right)$ & $\chi^2/dof$ \\*[1.0ex]
\hline\\*[-2ex]
\multirow{2}{*}{$\alpha=0.6$}
  & $5.8$ & $2.61^{+0}_{-12}$ & $2.01$ & $0.292^{+24}_{-4}$  & $15.7$ \\
  & $6.0$ & $1.11^{+5}_{-4}$  & $1.38$ & $0.996^{+57}_{-76}$ & $3.98$ \\*[1.0ex]

\multirow{2}{*}{$\alpha=0.5$}
  & $5.8$ & $1.89^{+0}_{-7}$  & $2.01$ & $0.487^{+38}_{-5}$  & $11.0$ \\
  & $6.0$ & $1.09^{+7}_{-6}$  & $1.38$ & $1.02^{+11}_{-10}$  & $5.56$ \\*[0.5ex]
\hline\hline
\end{tabular}
\label{tab:table6}
\end{table}
\begin{table}
\caption{Matching the ghost propagator for four $\beta$ values on
  $32^4$ lattices (Osaka data): shown are ratios of lattice spacings
  obtained by the matching relative to the finest one, for comparison
  the ratios predicted by Necco-Sommer scaling, the ratios of the
  renormalization constants and the $\chi^2/dof$ of the fit that
  accomplishes the matching, without ($\alpha=2.0$) and for two choices
  of the $\alpha$-cut.}
\centering
\begin{tabular}{c@{\quad}c@{\quad}c@{\quad}c@{\quad}c@{\quad}c}
\hline\hline\\*[-2.0ex]
$|p_i a| \le \alpha$ & $\beta$ & $\frac{a(\beta)}{a(6.1)}$ & 
$\frac{a^{\mathrm{NS}}(\beta)}{a^{\mathrm{NS}}(6.1)}$ &
$R_Z\left(\frac{a(\beta)}{a(6.1)}\right)$ & $\chi^2/dof$ \\*[1.0ex]
\hline\\*[-2ex]
\multirow{3}{*}{$\alpha=2.0$}
  & $5.8$ & $1.67^{+2}_{-2}$   & $1.73$ & $0.955^{+2}_{-3}$   & $1.63$  \\
  & $5.9$ & $1.39^{+2}_{-3}$   & $1.42$ & $0.971^{+3}_{-2}$   & $0.840$ \\
  & $6.0$ & $1.18^{+1}_{-2}$   & $1.18$ & $0.985^{+2}_{-1}$   & $0.260$ \\*[1.0ex]

\multirow{3}{*}{$\alpha=0.6$}
  & $5.8$ & $1.48^{+17}_{-11}$ & $1.73$ & $0.998^{+31}_{-44}$ & $0.219$ \\
  & $5.9$ & $1.34^{+9}_{-8}$   & $1.42$ & $0.981^{+23}_{-28}$ & $0.583$ \\
  & $6.0$ & $1.17^{+6}_{-6}$   & $1.18$ & $0.987^{+17}_{-17}$ & $0.420$ \\*[1.0ex]

\multirow{3}{*}{$\alpha=0.5$}
  & $5.8$ & $1.43^{+22}_{-21}$ & $1.73$ & $1.02^{+9}_{-7}$    & $0.268$ \\
  & $5.9$ & $1.35^{+17}_{-17}$ & $1.42$ & $0.979^{+61}_{-50}$  & $0.854$ \\
  & $6.0$ & $1.18^{+10}_{-8}$  & $1.18$ & $0.982^{+31}_{-29}$ & $0.501$ \\*[0.5ex]
\hline\hline
\end{tabular}
\label{tab:table7}
\end{table}
\begin{table}
\caption{Matching the ghost propagator for three $\beta$ values on
  $32^4$ lattices (Berlin data): shown are ratios of lattice spacings
  obtained by the matching relative to the finest one, for comparison
  the ratios predicted by Necco-Sommer scaling, the ratios of the
  renormalization constants and the $\chi^2/dof$ of the fit that
  accomplishes the matching, without ($\alpha=2.0$) and for two choices
  of the $\alpha$-cut.}
\centering
\begin{tabular}{c@{\quad}c@{\quad}c@{\quad}c@{\quad}c@{\quad}c}
\hline\hline\\*[-2.2ex]
$|p_i a| \le \alpha$ & $\beta$ & $\frac{a(\beta)}{a(6.2)}$ & 
$\frac{a^{\mathrm{NS}}(\beta)}{a^{\mathrm{NS}}(6.2)}$ &
$R_Z\left(\frac{a(\beta)}{a(6.2)}\right)$ & $\chi^2/dof$ \\*[1.0ex]
\hline\\*[-2ex]
\multirow{2}{*}{$\alpha=2.0$}
  & $5.8$ & $1.91^{+1}_{-1}$ & $2.01$ & $0.944^{+1}_{-1}$ & $10.4$ \\
  & $6.0$ & $1.34^{+1}_{-1}$ & $1.38$ & $0.976^{+1}_{-1}$ & $54.6$ \\*[1.0ex]

\multirow{2}{*}{$\alpha=0.6$}
  & $5.8$ & $1.80^{+1}_{-0}$ & $2.01$ & $0.959^{+1}_{-3}$ & $38.5$ \\
  & $6.0$ & $1.27^{+1}_{-1}$ & $1.38$ & $0.987^{+3}_{-2}$ & $121$  \\*[1.0ex]

\multirow{2}{*}{$\alpha=0.5$}
  & $5.8$ & $1.40^{+3}_{-3}$ & $2.01$ & $1.08^{+1}_{-1}$  & $18.5$ \\
  & $6.0$ & $1.18^{+2}_{-1}$ & $1.38$ & $1.02^{+1}_{-1}$  & $19.3$ \\*[0.5ex]
\hline\hline
\end{tabular}
\label{tab:table8}
\end{table}
\begin{table}
\centering
\caption{Fitted UV parameters and $\chi^2/dof$ for the transverse gluon 
propagator. Data from RCNP Osaka. $a=a(\beta=6.2)=0.1354 r_0$}
\centering
\begin{tabular}{c@{\quad}c@{\quad}c@{\quad}c@{\quad}c}
\hline\hline\\*[-2.0ex]
$|p_i a| \le \alpha$ & $[pa]_{\mathrm{min}}$ & 
$c_{\rm tr} a$ & $\eta_{\rm tr}$ & $\chi^2/dof$ \\*[1.0ex]
\hline\\*[-2ex]
\multirow{2}{*}{$\alpha=0.5$}
  & $0.5$ & $0.507(5)$  & $0.39(2)$ & $0.50$ \\
  & $0.6$ & $0.518(20)$ & $0.42(6)$ & $0.98$ \\*[1.0ex]

\multirow{4}{*}{$\alpha=0.6$}
  & $0.5$ & $0.534(5)$  & $0.46(1)$ & $0.81$ \\
  & $0.6$ & $0.537(6)$  & $0.46(2)$ & $0.70$ \\
  & $0.7$ & $0.517(15)$ & $0.43(3)$ & $0.75$ \\
  & $0.8$ & $0.489(50)$ & $0.39(6)$ & $1.55$  \\*[0.5ex]
\hline\hline
\end{tabular}
\label{tab:table9}
\end{table}
\begin{table}
\caption{Fitted UV parameters and $\chi^2/dof$ for 
the time-time dressing function $Z_{44}$ with
an $\alpha-$ cut for $\alpha = 0.5$. $a=a(\beta=6.0)=0.1863 r_0$.}
\centering
\begin{tabular}{c@{\quad}c@{\quad}c@{\quad}c@{\quad}c}
\hline\hline\\*[-2.0ex]
data & $[pa]_{\textrm{min}}$ & $c_{44} a$ & $\eta_{44}$ & $\chi^2/dof$ \\*[1.0ex]
\hline\\*[-2ex]
\multirow{4}{*}{Osaka}
 & $0.65$ & $0.942(8)$ & $2.53(10)$ & $1.75$  \\
 & $0.60$ & $0.946(6)$ & $2.47(6)$ & $1.19$  \\
 & $0.55$ & $0.945(4)$ & $2.48(4)$ & $0.81$  \\
 & $0.50$ & $0.940(4)$ & $2.54(3)$ & $2.18$  \\*[1.0ex]
\hline\\*[-2ex]
\multirow{5}{*}{Berlin}
 & $0.79$ & $0.968(16)$ & $2.09(22)$ & $0.23$ \\
 & $0.75$ & $0.962(10)$ & $2.19(11)$ & $0.26$  \\
 & $0.70$ & $0.962(9)$ & $2.19(10)$ & $0.18$  \\
 & $0.65$ & $0.948(5)$ & $2.38(5)$ & $1.16$  \\
 & $0.60$ & $0.940(4)$ & $2.49(4)$ & $2.22$  \\*[1.0ex]
\hline\hline
\end{tabular}
\label{tab:table10}
\end{table}
\begin{table}
\caption{Fitted IR parameters and $\chi^2/dof$ for the
time-time dressing function $Z_{44}$ with
an $\alpha-$ cut for $\alpha = 0.5$. $a=a(\beta=6.0)=0.1863 r_0$.}
\centering
\begin{tabular}{c@{\quad}c@{\quad}c@{\quad}c@{\quad}c}
\hline\hline\\*[-2.0ex]
data & $[pa]_{\textrm{max}}$ & $d_{44} a$ & $\kappa_{44}$ & $\chi^2/dof$ \\*[1.0ex]
\hline\\*[-2ex]
\multirow{3}{*}{Osaka}
 & $0.20$ & $2.07(14)$ & $1.71(5)$ & $0.057$  \\
 & $0.25$ & $1.52(4)$ & $1.94(3)$ & $19.5$  \\
 & $0.30$ & $1.36(3)$ & $2.05(2)$ & $20.3$  \\*[1.0ex]
\hline\\*[-2ex]
\multirow{3}{*}{Berlin}
 & $0.20$ & $4.25(32)$ & $1.26(3)$ & $18.2$ \\
 & $0.25$ & $2.75(8)$ & $1.44(2)$ & $21.2$  \\
 & $0.30$ & $2.16(4)$ & $1.57(1)$ & $47.0$  \\*[1.0ex]
\hline\hline
\end{tabular}
\label{tab:table11}
\end{table}
\begin{table}
\caption{Fitted UV parameters and $\chi^2/dof$ for the ghost
  dressing function ($a=a(\beta=6.0)=0.1863 r_0$.}
\centering
\begin{tabular}{c@{\quad}c@{\quad}c@{\quad}c@{\quad}c@{\quad}c}
\hline\hline\\*[-2.0ex]
  $|p_i a| \le \alpha$ & $[pa]_{\mathrm{min}}$ & $c_{\rm gh}$ &
  $\Lambda_{\rm{Coul}} a$ & $\gamma$ & $\chi^2/dof$ \\*[1.0ex] 
\hline\\*[-2ex]
\multirow{2}{*}{$\alpha=0.5$}
  & $0.70$ & $1.39(1)$   & $0.52(27)$  & $0.13(19)$ & $0.52$ \\
  & $0.75$ & $1.62(91)$  & $0.26(38)$  & $0.38(53)$ & $0.91$ \\*[1.0ex]

\multirow{2}{*}{$\alpha=0.6$}
  & $0.85$ & $1.58(6)$   & $0.277(28)$   & $0.365(28)$  & $3.88$  \\
  & $0.9$  & $1.44(9)$   & $0.370(81)$   & $0.289(58)$  & $2.81$  \\*[1.0ex]

\multirow{5}{*}{$\alpha=2.0$}
  & $2.0$  & $1.50(1)$   & $0.313(2)$    & $0.313(1)$   & $3.96$  \\
  & $2.5$  & $1.51(1)$   & $0.300(5)$    & $0.319(2)$   & $2.61$  \\
  & $3.0$  & $1.56(2)$   & $0.274(12)$   & $0.329(6)$   & $1.53$  \\
  & $3.5$  & $1.55(4)$   & $0.273(31)$   & $0.330(14)$  & $1.18$  \\
  & $4.0$  & $1.55(3)$   & $0.270(23)$   & $0.332(10)$  & $0.87$ \\*[0.5ex]
\hline\hline
\end{tabular}
\label{tab:table12}
\end{table}
\begin{table}
\caption{Fitted IR parameters and $\chi^2/dof$ for the ghost
    dressing function ($a=a(\beta=6.0)=0.1863 r_0$.}
\centering
\begin{tabular}{c@{\quad}c@{\quad}c@{\quad}c@{\quad}c}
\hline\hline\\*[-2.0ex]
$|p_i a| \le \alpha$ & $[pa]_{\mathrm{max}}$ & $d_{\rm gh} a$ & 
$\kappa_{\rm gh}$ & $\chi^2/dof$ \\*[1.0ex]
\hline\\*[-2ex]
\multirow{3}{*}{$\alpha=0.5$}
 & $0.29$ & $2.48(10)$ & $0.439(7)$ & $10.9$  \\
 & $0.30$ & $2.45(9)$ & $0.442(7)$ & $8.55$  \\
 & $0.34$ & $2.19(5)$ & $0.463(5)$ & $8.20$  \\*[1.0ex]

\multirow{3}{*}{$\alpha=0.6$}
 & $0.25$ & $2.57(10)$ & $0.435(6)$ & $0.62$ \\
 & $0.27$ & $2.54(9)$ & $0.437(6)$ & $3.79$  \\
 & $0.30$ & $2.50(9)$ & $0.440(6)$ & $4.32$  \\*[1.0ex]

\multirow{2}{*}{$\alpha=2.0$}
 & $0.24$ & $2.57(9)$ & $0.434(6)$ & $5.64$  \\
 & $0.30$ & $2.52(9)$ & $0.437(6)$ & $5.90$  \\*[0.5ex]
\hline\hline
\end{tabular}
\label{tab:table13}
\end{table}

\end{appendix}

\clearpage

\bibliographystyle{apsrev} 
\bibliography{citations_irqcd}

\end{document}